\documentclass[aps,prl,twocolumn,10pt]{revtex4-2}
\usepackage{amssymb}
\usepackage{graphicx}
\usepackage{amsmath}
\usepackage{hyperref}
\usepackage{subfigure}
\usepackage{multirow}
\usepackage{setspace}
\usepackage{verbatim}
\usepackage{float}
\usepackage{color}
\usepackage{ulem}
\usepackage[utf8]{inputenc}

\newcommand{\md}{\mathrm{d}}

\begin{document}

\title{Hydrodynamic sound shell model}

\author{Rong-Gen Cai$^{1,2,4}$}
\email{cairg@itp.ac.cn}

\author{Shao-Jiang Wang$^{2}$}
\email{schwang@itp.ac.cn}

\author{Zi-Yan Yuwen$^{2,3}$}
\email{yuwenziyan@itp.ac.cn (corresponding author)}

\affiliation{$^1$School of Physical Science and Technology, Ningbo University, Ningbo, 315211, China }
\affiliation{$^2$CAS Key Laboratory of Theoretical Physics, Institute of Theoretical Physics, Chinese Academy of Sciences, Beijing 100190, China}
\affiliation{$^3$School of Physical Sciences, University of Chinese Academy of Sciences, Beijing 100049, China}
\affiliation{$^4$School of Fundamental Physics and Mathematical Sciences, Hangzhou Institute for Advanced Study, University of Chinese Academy of Sciences, Hangzhou 310024, China}

\begin{abstract}
For a cosmological first-order phase transition in the early Universe, the associated stochastic gravitational wave background is usually dominated by sound waves from plasma fluid motions, which have been analytically modeled as a random superposition of freely propagating sound shells but with the force by the scalar field that produces the self-similar profile removed. In this Letter, we propose a new analytic sound shell model by focusing on the forced propagating contribution from the initial collision stage of sound shells when their self-similar profiles are still maintained by the moving bubble walls. We reproduce the causal $k^3$ scaling in the infrared consistent with numerical simulations, and also recover the broad dome in the power spectrum first observed in numerical simulations. The total sound waves should contain both contributions from forced collisions and free propagation of sound shells at early and late stages of the phase transition, respectively.
\end{abstract}
\maketitle

\textit{\textbf{Introduction.}---} 
The cosmological first-order phase transition (FOPT)~\cite{Mazumdar:2018dfl,Hindmarsh:2020hop,Caldwell:2022qsj}, if it exists, is a violent process in the early Universe, inducing large curvature perturbations~\cite{Liu:2022lvz} or even the formation of primordial black holes~\cite{Liu:2021svg} due to the asynchronous nature of vacuum-decay progress. The associated stochastic gravitational wave backgrounds (SGWBs)~\cite{Caprini:2015zlo,Caprini:2019egz} also open a new window into the early Universe that is otherwise opaque to light for us to probe the new physics~\cite{Cai:2017cbj,Bian:2021ini}  beyond the standard model of particles physics. The nonequilibrium feature also aids the realization for the baryon asymmetry~\cite{Cohen:1993nk,Morrissey:2012db} and primordial magnetic fields~\cite{Vachaspati:1991nm,Di:2020kbw,Yang:2021uid}.

The main sources of the SGWBs from cosmological FOPTs are bubble-wall collisions~\cite{Witten:1984rs,Hogan:1986qda} and bulk fluid motions from both sound waves~\cite{Hogan:1986qda} and magnetohydrodynamic turbulences~\cite{Witten:1984rs,Kamionkowski:1993fg}. Early numerical simulations~\cite{Kosowsky:1991ua,Kosowsky:1992rz,Kosowsky:1992vn,Kamionkowski:1993fg,Huber:2008hg} and analytical estimations~\cite{Caprini:2007xq,Caprini:2009fx} for the bubble-wall collisions have long adopted the so-called envelope approximation that dumps the wall upon collisions, which was abandoned in later numerical simulations~\cite{Hindmarsh:2013xza,Hindmarsh:2015qta,Hindmarsh:2017gnf} with thermal fluids, leading to the recognition of longitudinal acoustic waves as the dominant contribution~\cite{Weir:2016tov} as long-standing sources until the onset of vortical turbulences estimated both analytically~\cite{Kosowsky:2001xp,Dolgov:2002ra,Nicolis:2003tg,Caprini:2006jb,Gogoberidze:2007an,Caprini:2009yp} and numerically~\cite{Niksa:2018ofa,Pol:2019yex,Brandenburg:2021tmp,Brandenburg:2021bvg}.

For a vacuum phase transition without thermal fluids, the shape of the GW spectrum from bubble-wall collisions has been analytically modeled in Ref.~\cite{Jinno:2016vai} as a broken power law $(k^3, k^{-1})$ in the infrared (IR) and ultraviolet (UV) limits of the wave number $k$, respectively, by assuming thin-wall and envelope approximations, relaxing the latter of which analytically leads to the appearance of an intermediate linear growth~\cite{Jinno:2017fby} in addition to the original broken power law as $(k^3, k, k^{-1})$. This intermediate linear scaling is later confirmed in a semianalytical simulation from a dubbed bulk flow model~\cite{Konstandin:2017sat} beyond the envelope approximation but produces rather different UV behaviors as $(k,k^{-2})$ and $(k,k^{-3})$ for ultrarelativistic and nonrelativistic walls, respectively. Relaxing the thin-wall approximation would suppress the UV power to be steeper than $k^{-1}$ as found in the numerical simulation~\cite{Cutting:2020nla}. An additional peak seems to emerge in the UV regime close to the bubble-wall thickness and has been observed in the numerical simulation~\cite{Cutting:2018tjt} with a plausible explanation as scalar field oscillations around the true vacuum after vacuum decay.

On the other hand, for a thermal phase transition with plasma fluids, the shape of the GW spectrum is more involved as the contributions from bubble-wall collisions and bulk fluid motions are all mixed together. By detaching the fluid motions from the wall motion, a recently proposed sound shell model~\cite{Hindmarsh:2016lnk} has assumed freely propagating sound shells that were initially formed as self-similar profiles by hydrodynamics around the bubble wall, and later induces the fluid velocity field as a linear random superposition of an individual disturbance from each bubble. This sound shell model reveals the spectrum shape from sound waves as $(k^5,k,k^{-3})$ in the IR, intermediate, and UV regimes, respectively. The IR power $k^5$ was later corrected as $k^9$ in Ref.~\cite{Hindmarsh:2019phv} due to the causality for the divergence-free fluid velocity field~\cite{Durrer:2003ja}. However, both the numerical simulations with spectrum shape $(k^3,k^{-3})$~\cite{Hindmarsh:2013xza,Hindmarsh:2015qta,Hindmarsh:2017gnf} and a general theoretical expectation~\cite{Cai:2019cdl} prefer the usual $k^3$ scaling at low frequencies.

In this Letter, we propose a new analytic sound shell model by considering the initial collision stage when sound shells are still driven by the uncollided envelope of bubble walls. These forced propagating sound shells naturally lead to the usual causal $k^3$ scaling at low frequencies consistent with numerical simulations. We sketch the main assumptions and numerical fittings below and provide technical details in the Supplemental Appendix. 

\textit{\textbf{First-order phase transition.}---}
Depending on the underlying particle physics model with a FOPT~\cite{Kobakhidze:2017mru,Cai:2017tmh}, the nucleation rate can either exponentially increase with time or admit a local maximum value at some time~\cite{Jinno:2017ixd}. For the latter case dubbed the simultaneous nucleation~\cite{Cutting:2019zws,Cutting:2020nla}, the FOPT can never be ended if the maximal number density of bubbles ever nucleated is too small for percolation to be completed within one Hubble time~\cite{Guth:1982pn,Turner:1992tz}. Thus, the background expansion should be carefully accounted for in the case of simultaneous nucleation with general parameter choices~\cite{Ellis:2018mja}. Hence, for the sake of simplicity without considering the Hubble expansion effect, we will focus on the former case with an exponential nucleation rate (the number of nucleated bubbles per unit time and unit volume) of the form~\cite{Coleman:1977py, Callan:1977pt,Linde:1980tt, Linde:1981zj}
\begin{align} \label{eq:vacuum_decay_rate}
\Gamma(t) = \Gamma(t_*) e^{\beta(t - t_*)},
\end{align}
where $t_*$ is a fixed reference time usually chosen around bubble percolations, and $\beta^{-1}$ is roughly the time duration of the FOPT assumed here to be shorter than the Hubble time by $\beta/H\gg1$ so that the background Hubble expansion can be safely neglected. The strength factor $\alpha$ depicts the released latent heat of vacuum decay with respect to the background radiation energy density. The last parameter is the terminal wall velocity $v_w$ assumed here to be reached long before bubble collisions~\cite{Ellis:2019oqb,Ellis:2020nnr,Cai:2020djd,Lewicki:2022pdb}. As the SGWB from a FOPT is of most observational interest for a larger $v_w$, we will mainly focus on the detonation mode of bubble expansion with its supersonic terminal wall velocity larger than the Jouguet velocity~\cite{Steinhardt:1981ct}.

\textit{\textbf{Initial sound shell profile.}---} 
After bubble nucleations but before bubble collisions, we assume a steady expansion of spherical thin walls with a terminal velocity $v_w$. Since there is no characteristic scale during bubble expansion as the initial size of nucleated bubbles can be neglected, the wall expansion and associated fluid motions can be well described with spherical coordinates $(t,r,\theta,\varphi)$ by a single self-similar coordinate $\xi\equiv r/(t-t_n)$ tracing the fluid element at radius $r$ to the bubble center and at time $t$ since bubble nucleation time $t_n$. Without going into the second-order hydrodynamics with shear and bulk viscosity, the total energy-momentum tensor of the scalar-bubble/plasma-fluid system with a FOPT can be well approximated as a perfect fluid form~\cite{Wang:2022txy} $\hat{T}_{\mu\nu}=(\rho+p)U_\mu U_\nu+pg_{\mu\nu}$, with $\rho$, $p$, and $U^\mu$ denoting the energy density, pressure, and 4-velocity of bulk fluid, respectively. The conservation of $\hat{T}_{\mu\nu}$ further gives rise to the fluid equation of motion, which, under an additional assumption from a bag equation of state, can be explicitly solved for the fluid velocity profile~\cite{Espinosa:2010hh} given the junction conditions at the bubble wall and shockwave front (if any). The nonvanishing part of the fluid velocity profile will be referred to as the sound shell, consisting of compression and/or rarefaction waves of bulk fluids driven by the expanding bubble wall at least before the bubble collisions.

For the detonation mode of an expanding wall with $v_w$, the sound shell can be numerically solved as a rarefaction wave just behind the wall with the fluid velocity profile monotonically growing from zero at $\xi=c_s$ to a maximum value $v_m$ at $\xi=v_w$. For later convenience in analytic evaluations, we will adopt an analytical approximation
\begin{align}
v_i(t,\vec{x}) = v n_i \simeq \left\{ 
\begin{aligned}
&\frac{v_m(r - R_1(t))}{R_2(t) - R_1(t)} n_i,  & R_1(t) < r < & R_2(t)\\  
& 0, &\mathrm{otherwise}&
\end{aligned}
\right. \label{eq:vofr}
\end{align}
for the fluid velocity at a radial distance $r$ to the bubble center and time elapse $t-t_n$ since the nucleation time $t_n$. Here, $n_i = v_i / v$ is a unit vector from the bubble center toward the point $\vec{x}$, $v_m$ is the fluid velocity just behind the wall, and $R_1 = c_s(t - t_n)$ and $R_2 = v_w(t - t_n)$ are the innermost and outermost radii of the sound shell. This approximation can be obtained by replacing the curve between $R_1$ and $R_2$ with a straight line.

\begin{figure}
\centering
\includegraphics[width = 0.45\textwidth]{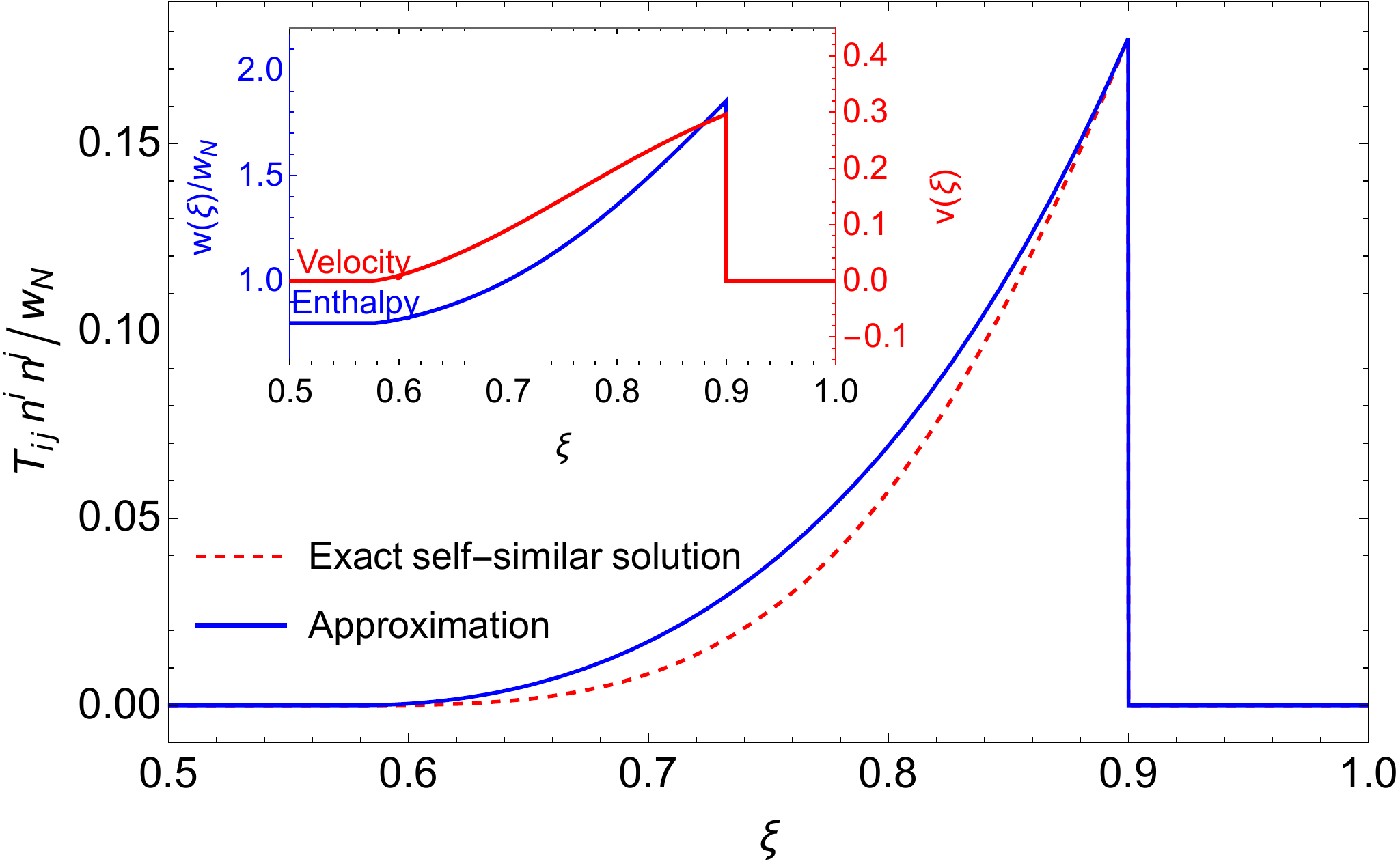}
\caption{The initial energy-momentum tensor profile from our approximation \eqref{eq:stress_tensor} compared to the exact numerical profile from the self-similar solution as a function of $\xi=r/(t-t_n)$ with $v_w=0.9$, $\alpha=0.2$. The exact numerical profiles of the fluid velocity and enthalpy are also presented in the inset.}
\label{fig:Tij_profiles}
\end{figure}

\textit{\textbf{Initial energy-momentum tensor profile.}---} 
For the computation of GWs, only the anisotropic spatial part of the energy-momentum tensor matters, that is, $T_{ij} = w \gamma^2 v_i v_j$, where $w = \rho + p$ is the enthalpy and $\gamma = (1-v^2)^{-1/2}$ is the Lorentz factor of the $3$-velocity $v_i$. Similar to the approximated velocity profile above, we further approximate the enthalpy profile also with a piecewise linear function as
\begin{align}
\frac{w(t,r)}{w_N} \simeq \left\{ 
\begin{aligned}
&\frac{w_r(r - R_1(t))}{R_2(t) - R_1(t)} + 1,  & R_1(t) < r < &R_2(t) \\  
&1,  &\mathrm{otherwise}&
\end{aligned}
\right. \label{eq:wbywnofr}
\end{align}
with $w_r = w_m / w_N - 1$, where $w_m$ and $w_N$ are the enthalpies just behind the bubble wall and at null infinity $\xi=1$, respectively. Note that as the enthalpy behind the sound shell $w(\xi< c_s)$ deviates a constant value from that far in front of the bubble wall $w_N$, and the contribution of the enthalpy term to the energy-momentum tensor is suppressed by the velocity as $\xi \to c_s$; hence, the ratio $w(\xi<c_s)/w_N$ can be approximated to be $1$. With approximations~\eqref{eq:vofr} and \eqref{eq:wbywnofr}, the initial energy-momentum tensor admits nonvanishing values only within the sound shell $R_1<r<R_2$ as
\begin{align}
T_{ij} = & \frac{w v^2}{1-v^2} n_i n_j \nonumber \\
= & w_N \left( \frac{w_r}{v_w - c_s}\frac{r - c_s (t - t_n)}{t - t_n} + 1 \right)  \nonumber \\ 
& \quad \times \sum_{s = 0}^{\infty} \left( \frac{v_m}{v_w - c_s}\frac{r - c_s (t - t_n)}{t - t_n} \right)^{2 s + 2} n_i n_j,
\label{eq:stress_tensor}
\end{align}
where the expansion is sufficient to take the first three terms $s=0,1,2$, as the maximal bulk fluid velocity $v_m$ is of order $\mathcal{O}(10^{-1})$. The comparison of $T_{ij}n^i n^j = w\gamma^2 v^2$ between our analytic approximation and the exact numerical evaluation is shown in Fig.~\ref{fig:Tij_profiles}, along with which the profiles of the  enthalpy and fluid velocity from the exact self-similar solution are also shown in the inset. This energy-momentum tensor will maintain its initial profile~\eqref{eq:stress_tensor} until the driven walls collide with each other, after which, the part of sound shells still driven by the uncollided envelopes of walls will continue to keep its initial hydrodynamic profile, while the remaining part of the sound shells will propagate freely with damped amplitude and widened thickness as investigated in Ref.~\cite{Hindmarsh:2016lnk}. In the rest of this Letter, we will focus on the former contribution from the forced propagating sound shells that is usually overlooked in the literature.

\begin{figure}
\centering
\includegraphics[width = 0.35\textwidth]{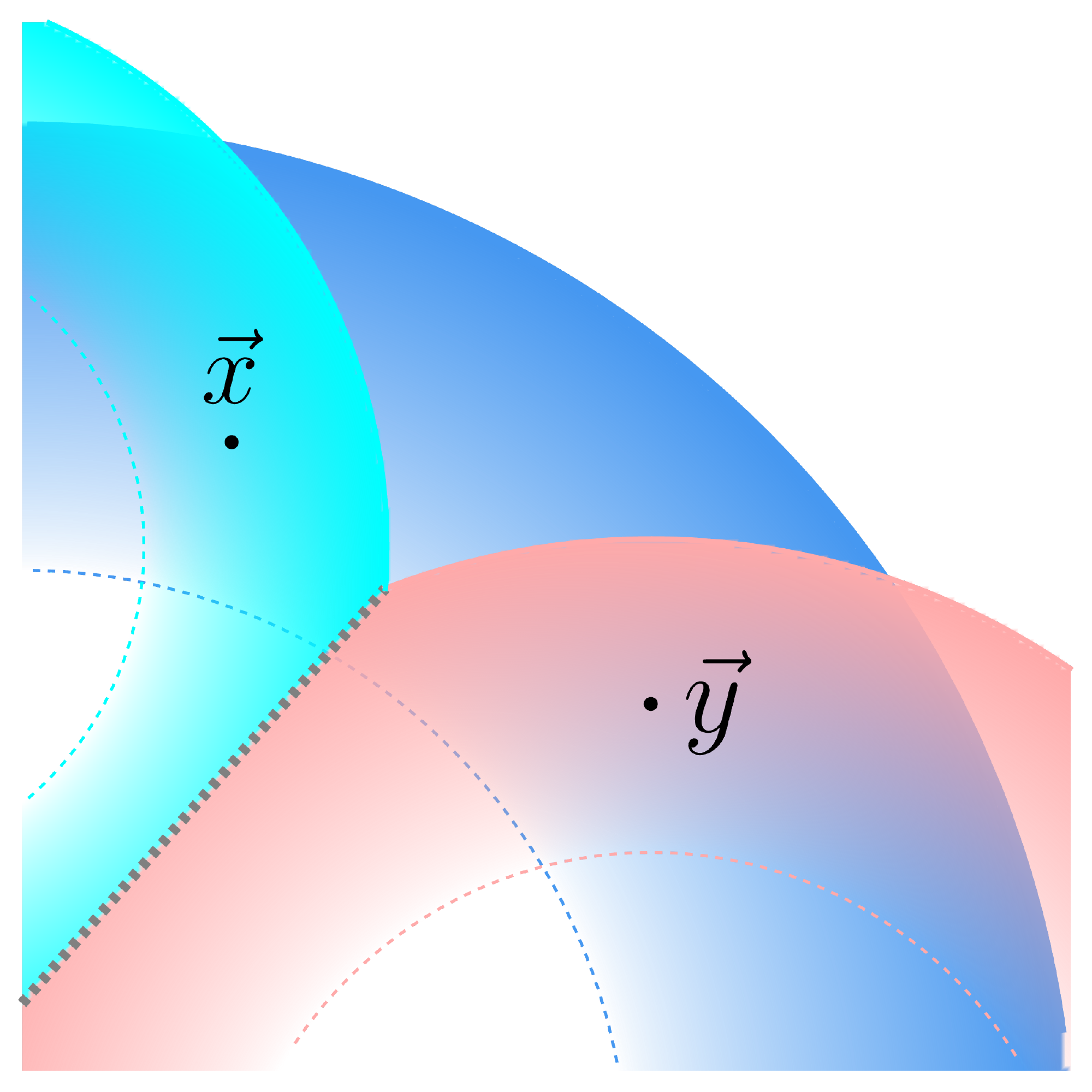}
\caption{A schematic illustration of the single-shell and double-shell contributions to the energy-momentum tensor with their two-point correlation functions coming from $\vec{x}$ and $\vec{y}$ in the same sound shell (blue shell) or in different sound shells (cyan and red shells).}
\label{fig:single_double_bubble}
\end{figure}

\begin{figure}
    \centering
    \includegraphics[width = 0.48\textwidth]{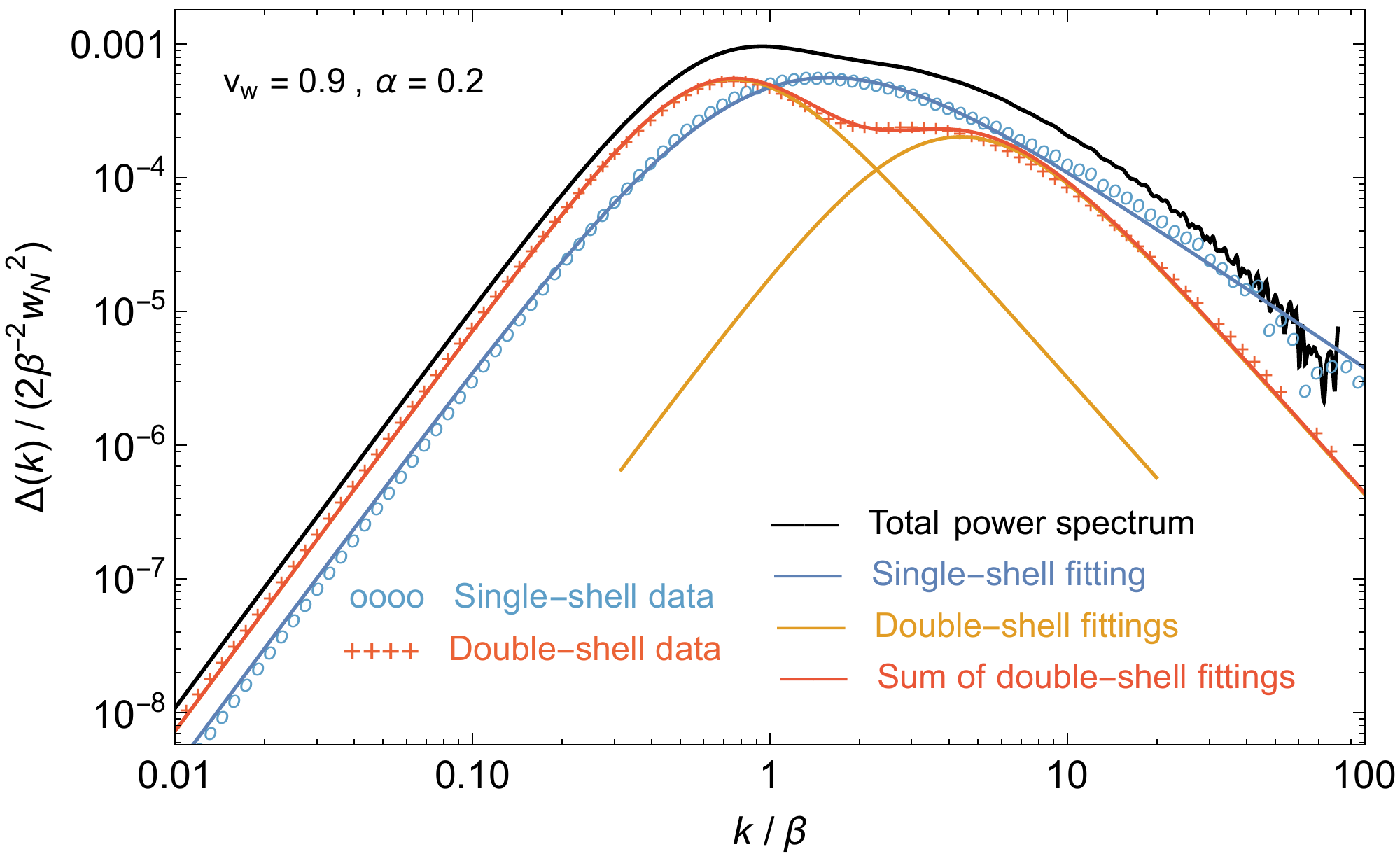}
    \caption{The dimensionless power spectrum (black solid) from single-shell (blue circles) and double-shell (red crosses) contributions fitted by single broken-power-law (blue solid) and double broken-power-law (orange and red solid) templates, respectively, for illustrative parameters $v_w=0.9$ and $\alpha = 0.2$.}
    \label{fig:Total_Power}
\end{figure}

\textit{\textbf{Sound shell forced collisions.}---} 
Forced collisions of sound shells driven by uncollided envelopes of bubble walls will generate a GW energy-density power spectrum $P_\mathrm{GW}(t,k)\equiv\mathrm{d}(\rho_\mathrm{GW}/\rho_\mathrm{tot})/\mathrm{d}\ln k\equiv 2Ga_*^4/(\pi\rho_\mathrm{tot}a^4)\Delta(k)$ from a two-point correlation function $\langle T_{ij}(x)T_{kl}(y) \rangle$ at two space-time points $x$ and $y$. Here, $\rho_\mathrm{GW}$ and $\rho_\mathrm{tot}$ are GW and critical energy densities at time $t$ with scale factor $a(t)$, respectively, redshifted from the dimensionless spectrum $\widetilde{\Delta}(k)\equiv\Delta(k)/(2\beta^{-2}w_N^2)$ at the phase-transition completion $t_*$ with scale factor $a(t_*)\equiv a_*$. When $x$ and $y$ are in the same (different) sound shell, the two-point correlator $\langle T_{ij}(x)T_{kl}(y) \rangle$ serves as the single-shell (double-shell) contribution as illustrated in the schematic picture of Fig.~\ref{fig:single_double_bubble}. After tedious and lengthy calculations as detailed in the Supplemental Appendix, 
the above two contributions can be analytically expressed as formal integrals, which can be further evaluated numerically given the wall velocity $v_w$ and strength factor $\alpha$. With a typical choice for the parameters $v_w=0.9$ and $\alpha=0.2$, the numerical integration results for the single-shell spectrum $\widetilde{\Delta}^{(s)}$ (red crosses) and double-shell spectrum $\widetilde{\Delta}^{(d)}$ (blue circles) to the total GW power spectrum $\widetilde{\Delta}=\widetilde{\Delta}^{(s)}+\widetilde{\Delta}^{(d)}$ (black solid) are shown in Fig.~\ref{fig:Total_Power} with numerical fittings.

For practical use in extracting phase-transition parameters from numerical simulations as well as future GW observations, we provide here the fitting template from proper combinations of the broken-power-law ansatz
\begin{align}
    F_{n_1,n_2,\delta}(k;k_*,F_*) = F_* \left(\frac{k}{k_*}\right)^{n_1} \left( \frac{1+(k/k_*)^{\delta}}{2} \right)^{\frac{n_2-n_1}{\delta}}
\end{align}
depicting a peak amplitude $F_*$ at the peak frequency $k_*$ with a peak transition width $\delta^{-1}$ from a low-frequency slope $n_1$ to a high-frequency slope $n_2$. For our illustrative example with $\alpha=0.2$ and $v_w=0.9$, the single-shell power spectrum $\widetilde{\Delta}^{(s)}$ is well fitted by this broken-power-law shape as
\begin{align}\label{eq:single_fit}
    \widetilde{\Delta}^{(s)}(k) = F_{n_1,n_2,\delta}(k;k_*,F_*)
\end{align}
with parameters $n_1=3$, $n_2=-3/2$, $\delta=3/2$, $k_*=\beta$, and $F_*=4.7\times 10^{-4}$, which asymptotes to $k^3$ at low frequencies and $k^{-3/2}$ at high frequencies. 
The double-shell power spectrum $\widetilde{\Delta}^{(d)}$ admits two peaks that can be well fitted by the sum of two broken-power-law shapes as
\begin{align}\label{eq:double_fit}
    \widetilde{\Delta}^{(d)}(k) = F_{n_1,n_2,\delta_1}(k;k_{*1},F_{*1}) + F_{n_1,n_2,\delta_2}(k;k_{*2},F_{*2})
\end{align}
with parameters $n_1 = 3$, $n_2 = -5/2$, $\delta_1 = 2$, $\delta_2 = 3/2$, $k_{*1}=0.7\beta$, $k_{*2}=4\beta$, $F_{*1}=5.4\times 10^{-4}$, and $F_{*2}=1.0\times 10^{-4}$, which again asymptotes to $k^3$ at low frequencies but $k^{-5/2}$ at high frequencies.

More general results with another fixed $\alpha=0.1$ but varying $v_w=0.8-1.0$ are shown in the Supplemental Appendix, 
from which we can learn that, for both contributions, the low-frequency behaviors always recover the causal $k^3$ scaling, consistent with both numerical simulations~\cite{Hindmarsh:2013xza,Hindmarsh:2015qta,Hindmarsh:2017gnf} and general analytic expectation~\cite{Cai:2019cdl}. Furthermore,
the double-shell contribution always dominates over the single-shell contribution at low frequencies, while the single-shell contribution would gradually take over the high-frequency dominance for an increasing wall velocity. 
Therefore, we reproduce a broader dome in the total power spectrum with a decreasing detonation wall velocity as first observed in numerical simulations~\cite{Hindmarsh:2013xza,Hindmarsh:2015qta,Hindmarsh:2017gnf}.
However, except for the universal $k^3$ scaling at low frequencies, all other parameters admit mild extra dependence on the bubble-wall velocity and phase-transition duration as summarized below.

\noindent\textbf{(i) \textit{Peak amplitudes.}---}
For the case attached in the Supplemental Appendix 
with fixed $\alpha=0.1$ but varying $v_w=0.8-1.0$, the peak amplitudes of single-shell and double-shell spectra can be naively fitted as
\begin{align}
    \widetilde{\Delta}_*^{(s)} &= \frac{1 -1.84 v_w + 0.85 v_w^2}{-4.40 + 5.50 v_w - 9.96 v_w^2} \times 10^{-4} , \\
    \widetilde{\Delta}_{*1}^{(d)} &= \frac{1 -1.89 v_w + 0.90 v_w^2}{-3.33 + 7.33 v_w - 3.90 v_w^2} \times 10^{-4} , \\
    \widetilde{\Delta}_{*2}^{(d)} &=  \frac{1 -1.96 v_w + 0.96 v_w^2}{-1.58 + 2.94 v_w + 1.14 v_w^2} \times 10^{-4}.
\end{align}
The peak amplitudes for other values of the strength factor can be related to the above example with $\alpha=0.1$ by a simple scaling relation since the pure $\alpha$ dependence can be factorized out approximately in a form like
\begin{align}\label{eq:Power_approx}
    \Delta(k,v_w,\alpha) \simeq \left(w_m(v_w,\alpha)v_m^2(v_w,\alpha)\gamma_m^2(v_w,\alpha)\right)^2 \hat{\Delta}(k,v_w),
\end{align}
where $w_m(v_w,\alpha)$ and $v_m(v_w,\alpha)$ can be analytically derived from the junction condition at the bubble wall~\cite{Espinosa:2010hh}. This is because, as the maximum fluid velocity $v_m$ is small and $w_r=w_m/w_N -1$ is also not large, the energy-momentum profiles of different $\alpha$ could be approximately related by
\begin{align}
    \frac{T_{ij}(v_w,\alpha_1) n^i n^j}{T_{kl}(v_w,\alpha_2) n^k n^l} = \frac{w_m(v_w,\alpha_1)v_m^2(v_w,\alpha_1)\gamma_m^2(v_w,\alpha_1)}{w_m(v_w,\alpha_2)v_m^2(v_w,\alpha_2)\gamma_m^2(v_w,\alpha_2)}.
\end{align}
Therefore, the peak frequencies and spectrum slope are essentially encoded in $\hat{\Delta}(k,v_w)$ independent of $\alpha$, while the peak amplitudes can be transformed back and forth by applying Eq.~\eqref{eq:Power_approx} as long as any one of them is known.

\noindent\textbf{(ii) \textit{Peak frequencies.}---} 
To physically fit the peak frequencies from our numerical results, we first identify two characteristic length scales. The first scale is the averaged separation of bubbles at the onset of nucleation, which is twice the averaged bubble radius $R_w = (8\pi)^{1/3} v_w \beta^{-1}\equiv k_w^{-1}$ at collisions. The second scale is the thickness of sound shell $L_s = R_w(v_w - c_s)/v_w\equiv k_s^{-1}$. Then, the peak frequency for the single-shell spectrum can be well fitted at 
\begin{align}\label{eq:single_peak_fitting}
    k_*^{(s)} = 3.78 k_w,
\end{align}
and for the double-shell spectrum, the lower peak frequency can be well fitted at
\begin{align}\label{eq:double_peak1_fitting}
    k_{*1}^{(d)} = C 
    \left( \frac{k_s}{k_w} \right)^n k_s = C \left(\frac{k_w}{k_s} \right)^m k_w
\end{align}
with $n = -0.74$, $m+n=-1$, and $C = 1.54$, while the higher peak frequency can be well fitted at
\begin{align}\label{eq:double_peak2_fitting}
    k_{*2}^{(d)} = \frac{1 - (2-\delta_1) v_w + (1+\delta_1)v_w^2}{\delta_2 + (0.1+\delta_3) v_w - (0.1-\delta_3)v_w^2}
\end{align}

\noindent\textbf{(iii) \textit{Spectrum slopes.}---}
Because of numerical errors from integrating irregular regions of sound shell collisions, spectrum slopes at high frequencies are extremely difficult to be determined precisely, especially the single-shell spectrum. For the example shown in the Supplemental Appendix, 
the high-frequency slopes of single-shell spectrum vary roughly from $k^{-2}$ to $k^{-1}$ as $v_w$ increasing from $0.80$ to $1.00$, while for the double-shell spectrum, the high-frequency slopes drop from $k^{-5/2}$ to $k^{-3}$ more confidently. These behaviors might be related to the sound-shell thickness, similar to that observed previously in the numerical simulation~\cite{Cutting:2020nla} of wall collisions with varying wall thicknesses. This allows us to constrain $v_w$ from spectrum slopes besides the usual peak frequencies and amplitudes if SGWBs of FOPTs are observed in future. It might as well be that the correct UV slope be captured by free collisions of sound shells. We leave the precise $v_w$ dependence of spectrum slopes in future.

\noindent\textbf{(iv) \textit{Present spectrum.}---} 
The SGWBs from FOPTs propagate as non-interacting radiations whose energy density evolves with $a^{-4}$. The scale factor $a_*$ at the end of phase transition is related to that $a_0$ at present by~\cite{Jinno:2016vai}
\begin{align}
    \frac{a_*}{a_0} = 8.0 \times 10^{-16} \left(\frac{g_*}{100}\right)^{-\frac{1}{3}} \left(\frac{T_*}{100 \mathrm{GeV}}\right)^{-1},
\end{align}
where $g_*$ is the total number of degrees of freedom of  relativistic species at phase-transition temperature $T_*$. The peak frequency at present redshifted by $a_*/a_0$ reads
\begin{align}
k_0 = 1.65\times 10^{-5} \mathrm{Hz} \left(\frac{g_*}{100}\right)^{\frac{1}{6}} \left(\frac{T_*}{100 \mathrm{GeV}}\right)\left( \frac{\beta}{H_*} \right)\left( \frac{k_*}{\beta} \right).
\end{align}
The GW power spectrum at present is redshifted as
\begin{align}
&\frac{\mathrm{d}\Omega_\mathrm{GW}}{\mathrm{d}\ln k}=1.67\times10^{-5}\left(\frac{g_*}{100}\right)^{-\frac13}\frac{2G}{\pi\rho_\mathrm{tot}}(2\beta^{-2}w_N^2)\widetilde{\Delta}(k)\nonumber\\
&=4.51\times10^{-6}\left(\frac{g_*}{100}\right)^{-\frac13}\left(\frac{H_*}{\beta}\right)^2\left(\frac{1}{1+\alpha}\right)^2\widetilde{\Delta}(k),
\end{align}
where the total energy density $3H_*^2/(8\pi G)=\rho_{\mathrm{tot}} = \rho_{\mathrm{rad}} + \epsilon$ consists of the released vacuum energy $\epsilon=\alpha\rho_\mathrm{rad}$ and thermal radiations with the enthalpy $w_N = (4/3) \rho_{\mathrm{rad}}$. Compared to the usual GW spectrum template with the combination $[\kappa_v\alpha/(1+\alpha)]^2$ factorized out, we prefer to keep the efficiency factor $\kappa_v$~\cite{Espinosa:2010hh} hidden in the dimensionless $\widetilde{\Delta}$ due to the scaling relation~\eqref{eq:Power_approx}.

\textbf{\textit{Conclusions and discussions.}---} 
The SGWBs from the bulk fluid motions, especially the sound waves, are the dominant GW sources for cosmological FOPTs provided that most of bubble walls collide with each other long after they have approached the terminal velocity~\cite{Cai:2020djd,Lewicki:2022pdb}. Because of the limited computational power, numerical simulations are usually implemented for a limited parameter space, making the analytic auxiliary modeling an indispensable tool to extract from numerical simulations the fitting templates of GW spectrum that  are used for specific model predictions. The analytic auxiliary modeling~\cite{Hindmarsh:2016lnk,Hindmarsh:2019phv} for GWs from sound waves has only considered the late-time free collisions of sound shells but overlooked the early-time forced collisions of sound shells, which has been computed analytically in the present work for the detonation mode of bulk fluid motions. We have successfully recovered the causal $k^3$ scaling at low frequencies and revealed the underlying structure of a widened dome around the peak frequency from a combination of single-shell and double-shell contributions, all consistent with numerical simulations. The final sound-wave spectrum $(k^3, k^{-3})$ suggested by numerical simulations should be a combination of the forced and free collisions of sound shells producing the $k^3$ scaling and $k^{-3}$ scaling in the IR and UV regimes, respectively,
\begin{align}
\frac{\mathrm{d}\Omega_\mathrm{GW}}{\mathrm{d}\ln k}\bigg|_\mathrm{SW}=\frac{\mathrm{d}\Omega_\mathrm{GW}}{\mathrm{d}\ln k}\bigg|_\mathrm{SW}^\mathrm{forced}+\frac{\mathrm{d}\Omega_\mathrm{GW}}{\mathrm{d}\ln k}\bigg|_\mathrm{SW}^\mathrm{free}.
\end{align}
Several improvements can be made in future works for better analytic auxiliary modelings as follows:

First, for completeness, the GWs from forced collisions of sound shells should also be computed for deflagration cases of Jouguet and weak types, the latter of which has been recently shown to be feasible for strongly coupled FOPTs~\cite{LiLi:2023dlc}. All these calculations should be carried out for more realistic sound shell profiles beyond the simple linear interpolation and bag equation of state~\cite{Giese:2020znk,Giese:2020rtr,Wang:2020nzm,Wang:2022lyd,Wang:2023jto}.

Second, our analytic sound waves from forced collisions of sound shells driven by the envelope of uncollided walls have neglected the contributions from the overlapping parts of colliding sound shells. This envelope approximation of sound-shell forced collisions might be the reason why our high-frequency slopes from both single-shell and double-shell spectra deviate from numerical simulations. 

Third, although our analytic model has achieved much better agreement with simulations than either sound shell model~\cite{Hindmarsh:2016lnk,Hindmarsh:2019phv} or bulk flow model~\cite{Jinno:2017fby,Konstandin:2017sat}  by reproducing the causal IR scaling and double-peak structure, respectively,  the envelope approximation we have adopted for the forced collisions of sound shells during the percolation stage, similar to the bulk flow model, also results in an extra suppression factor $H_*/\beta$ in the GW spectrum compared to the sound shell model and numerical simulation results. Future analytic study of sound shell collisions both during and after bubble percolations should go beyond the envelope approximation of sound shells.

Last, the Hubble expansion effect has been considered in previous modelings of wall collisions~\cite{Zhong:2021hgo} and sound waves~\cite{Guo:2020grp}, which should also be accounted for even though we expect it to be small as our forced collisions of sound shells are mainly important at the early stage of collisions when the sound shells still maintain their self-similar profiles.

\begin{acknowledgments}
\textbf{\textit{Acknowledgments.}---} This work is supported by the National Key Research and Development Program of China Grant  No. 2021YFC2203004, No. 2021YFA0718304, and No. 2020YFC2201502,
the National Natural Science Foundation of China Grants No. 12105344, No. 12235019, No. 11821505, No. 11991052, and No. 11947302,
and the Science Research Grants from the China Manned Space Project with No. CMS-CSST-2021-B01.
We acknowledge the use of HPC of ITP-CAS.
\end{acknowledgments}

\appendix

\section{Supplemental Appendix}

In this supplemental material, we present in details the analytic derivations on the single-shell and double-shell contributions to the total gravitational-wave energy-density power spectrum from the forced collisions of sound shells driven by the uncollided envelope of bubble walls during the percolation stage.

\section{Conventions}\label{sec:GW_power_spectrum}

We first closely follow the conventions of Refs.~\cite{Caprini:2007xq, Jinno:2016vai} to set up the notations for the gravitational-wave (GW) power spectrum. We consider a spatially flat Universe with tensor perturbations,
\begin{align}
    \md s^2 = a(t)^2(- \md t^2 + (\delta_{ij}+2h_{ij})\md x^i \md x^j),
\end{align}
where $t$ is the conformal time and $h_{ij}$ denotes the tensor perturbations in the transverse-traceless gauge with $h_{ii} = \partial_i h_{ij} = 0$. The Fourier transformation of the equation of motion for $h_{ij}$, to the linear order, reads
\begin{align}
    \partial_t^2 h_{ij}(t,\vec{k}) + 2\mathcal{H}\partial_t h_{ij}(t,\vec{k})& + k^2 h_{ij}(t,\vec{k}) \nonumber \\
    & = 8\pi G a(t)^2 \Pi_{ij}(t,\vec{k}),
    \label{eq:hEoM}
\end{align}
where $\mathcal{H} = \md \ln a / \md t$, $\Pi_{ij}(t,\vec{k}) = \Lambda_{ij,kl}(\hat{k})T_{kl}(t,\vec{k})$ is the Fourier transform of the transverse-traceless energy-momentum tensor, and $\Lambda_{ij,kl}(\hat{k})$ is the projection tensor defined by
\begin{align}
    \Lambda_{ij,kl}(\hat{k}) & = P_{ik}(\hat{k}) P_{jl}(\hat{k}) - \frac{1}{2} P_{ij}(\hat{k}) P_{kl}(\hat{k}), \\
    P_{ij}(\hat{k}) & = \delta_{ij} - \hat{k}_i \hat{k}_j.
\end{align}
We assume that the phase transition is completed within one Hubble time, hence the Hubble expansion effect could be neglected. Thus, we can drop the second term in the left hand side of \eqref{eq:hEoM} and then approximate $a(t)$ in the right hand side with its value at the transition time by $a_* \equiv a(t_*)$, where $t_*$ denotes the time of the phase transition completion. Hence, we can rewrite \eqref{eq:hEoM} as
\begin{align}
    \partial_t^2 h_{ij}(t,\vec{k}) + k^2 h_{ij}(t,\vec{k}) = 8\pi G a_*^2 \Pi_{ij}(t,\vec{k}).
    \label{eq:hEoMLinear}
\end{align}

The solutions to the above equation can be formally expressed with the Green function method. We assume the source term is active from $t_i$ to $t_f$, and in later calculations we will take the limit $t_{i,f}\to\mp\infty$. One might wonder whether the limit $t_i\to -\infty$ is reasonable, since $t_i$ as the conformal time should be $t_i > t_0=0$ for FOPTs occurring in the radiation dominated Universe, where $t_0$ is the conformal time at the end of inflation. Now that the time $t$ enters our later calculation with a factor of the form $\exp(\beta(t-t_*))$ exponentially suppressed as $t\to t_i$, therefore, we simply take the limit $t_i\to -\infty$ for later convenience in analytical integration. Then, for $t<t_f$, the solution to Eq.~\eqref{eq:hEoMLinear} is
\begin{align}
    h_{ij}(t,\vec{k}) = \frac{8\pi G a_*^2}{k} \int_{t_i}^{t_f} \md t' \mathcal{G}_k(t,t') \Pi_{ij}(t,\vec{k}),
\end{align}
where $\mathcal{G}_k(t,t') = \sin(k(t-t'))/k$ is the Green function satisfying $\mathcal{G}_k(t,t) = 0$, $\partial_t\mathcal{G}_k(t,t')|_{t=t'} = 1$. As for $t>t_f$, the source term vanishes. Thus, we have to match the solutions to free waves during radiation domination at $t_f$, giving rise to
\begin{align}
    h_{ij}(t,\vec{k}) = A_{ij}(\vec{k}) \frac{\sin(k(t-t_f))}{kt} + B_{ij}(\vec{k}) \frac{\cos(k(t-t_f))}{kt},
    \label{eq:solutions_to_EoM}
\end{align}
with coefficients
\begin{align}
    A_{ij}(\vec{k}) &= \frac{8\pi G a_*^2}{k} kt_f \int_{t_i}^{t_f}\md t \cos(k(t_f - t)) \Pi_{ij}(t,\vec{k}), \\
    B_{ij}(\vec{k}) &= \frac{8\pi G a_*^2}{k} kt_f \int_{t_i}^{t_f}\md t \sin(k(t_f - t)) \Pi_{ij}(t,\vec{k}).
\end{align}

We define the GW power spectrum by the equal-time correlator given by
\begin{align}
    \langle \partial_t h_{ij}(t,\vec{p}) \partial_t h_{ij}^*(t,\vec{q}) \rangle = (2\pi)^3 \delta^{(3)}(\vec{p} - \vec{q}) P_{\dot{h}}(t,p),
\end{align}
where the bracket $\langle\dots\rangle$ denotes the ensemble average. We also define the unequal-time correlator of the source term by
\begin{align}
    \langle \Pi_{ij}(t_x,\vec{p})\Pi_{ij}^*(t_y,\vec{q}) \rangle = (2\pi)^3 \delta^{(3)}(\vec{p} - \vec{q}) \Pi(t_x,t_y,p).
\end{align}
Using the solution Eq.~\eqref{eq:solutions_to_EoM}, we can express $P_{\dot{h}}$ in terms of the source terms. After dropping higher-order terms, the main contributions to the power spectrum from the source terms are
\begin{align}
    & P_{\dot{h}}(t,\vec{k})  \simeq \frac{1}{2 t^2} \left( \langle A_{ij}A_{ij}^* \rangle + \langle B_{ij}B_{ij}^*\rangle\right) \nonumber\\
    & \quad = \left( 8\pi G a_*^2 \right)^2 \frac{t_f^2}{2t^2} \nonumber\\
    & \quad \quad  \times \int_{t_i}^{t_f}\md t_x \int_{t_i}^{t_f} \md t_y \cos(k(t_x - t_y)) \Pi(t_x,t_y,k),
\end{align}
With $\vec{r} \equiv \vec{x} - \vec{y}$, the two-point correlator $\Pi(t_x,t_y,k)$ reads
\begin{align}
    \Pi(t_x,t_y,k) =& \Lambda_{ij,kl}(\hat{k}) \Lambda_{ij,mn}(\hat{k}) \nonumber \\
    & \quad \times \int \md^3 r e^{i \vec{k} \cdot \vec{r}} \langle T_{kl}(t_x,\vec{x})T_{mn}(t_y,\vec{y})\rangle.
    \label{eq:stress_correlator}
\end{align}

Now we can evaluate the energy density of GWs defined by
\begin{align}
    \rho_{\mathrm{GW}} & = \frac{\langle \partial_t h_{ij}(t,\vec{x}) \partial_t h_{ij}(t,\vec{x}) \rangle_T}{8\pi G a^2(t)} 
    \label{eq:rhoGW}
\end{align}
where $\langle\dots\rangle_T$ denotes the oscillation and ensemble average for a stochastic background. The power spectrum is defined by the energy density per logarithmic frequency and using the definition \eqref{eq:rhoGW} we can get
\begin{align}
    & P_{\mathrm{GW}}(t,k) \nonumber \equiv \frac{1}{\rho_\mathrm{tot}} \frac{\md \rho_\mathrm{GW}}{\md \ln k} \nonumber\\
    & \quad = \frac{2G}{\pi \rho_{\mathrm{tot}}} \frac{a_*^4}{a^4}  k^3 \nonumber \\
    & \quad \quad \times \int_{t_i}^{t_f}\md t_x \int_{t_i}^{t_f} \md t_y \cos(k(t_x - t_y)) \Pi(t_x,t_y,k),
    \label{eq:PGW}
\end{align}
where we have set $t_f \simeq t_*$ and replaced $t_f/t$ with $a_*/a$ during radiation domination. For later convenience, we define a time-independent power spectrum $\Delta$ as 
\begin{align}
    \Delta(k) \equiv P_{\mathrm{GW}}\bigg/ \left( \frac{2G}{\pi \rho_{\mathrm{tot}}} \frac{a_*^4}{a^4} \right).
\end{align}
This quantity describes the power spectrum at the phase-transition time, and its later time evolution is contained in $2Ga_*^4/(\pi \rho_\mathrm{tot}a^4)$. The main purpose in the next section is to evaluate $\Delta(k)$.

\section{Assumptions}\label{sec:derivation}

We next present the detailed analytical derivation of the GW power spectrum from forced collisions of sound shells. More precisely speaking, we only focus on GWs from forced propagating sound shells \textit{during bubble collisions}, not after the collisions. For sound waves after collisions, considerable efforts have been made by assuming a Gaussian distribution of the bulk fluid velocity in the momentum space\cite{Hindmarsh:2016lnk, Hindmarsh:2019phv, Guo:2020grp}. The main assumptions they made contain the following two aspects. First, the sound waves propagate freely after bubble collisions, which might be not acceptable if the interactions in plasma fluid are strong enough. Second, they assume a Gaussian distribution for the fluid velocity field, which might not be true if one takes the velocity profile into consideration.

The main task is to evaluate the two-point correlator of energy-momentum tensor Eq.~\eqref{eq:stress_correlator} in traceless-transverse gauge. For the sake of later convenience, we will call the hypersurface satisfying $t^2 - r^2/v^2 =0$ as the $v$-cone similar to the light cone with $v=c=1$. The $c_s$-cone will also be referred to as a sound cone, where $c_s$ is the speed of sound. Under a bag equation of state, we have $c_s = 1/\sqrt{3}$. Given two points $x = (t_x,\vec{x})$ and $y = (t_y,\vec{y})$, in order to obtain a non-zero correlator of uncollided energy-momentum tensor given in the main context,
\begin{align}
    T_{ij} = & \frac{w v^2}{1-v^2} n_i n_j \nonumber \\
    = & w_N \left( \frac{w_r}{v_w - c_s}\frac{r - c_s (t - t_n)}{t - t_n} + 1 \right)  \nonumber \\ 
     & \quad \times \sum_{s = 0}^{\infty} \left( \frac{v_m}{v_w - c_s}\frac{r - c_s (t - t_n)}{t - t_n} \right)^{2 s + 2} n_i n_j,
    \label{eq:stress_tensor_app}
\end{align}
the following two conditions should be satisfied:

\begin{enumerate}
    \item There should be only one sound shell passing through $x$ and only one sound shell passing through $y$ since we only consider the uncollided part such that Eq.~\eqref{eq:stress_tensor_app} can be applied. 
    \item There should be no bubble ever nucleated inside the past $v_w$-cones of $x$ and $y$. The front and back ends of the sound shell travel at the speeds of bubble wall $v_w$ and speed of sound $c_s$, respectively. Therefore, either $x$ or $y$ is already passed by the sound shell if any bubble could be nucleated inside the sound cone. This condition will be taken into account later through Eq.~\eqref{eq:Possibility_in_FV}.
\end{enumerate}

\begin{figure}
    \centering
    \includegraphics[width = 0.4\textwidth]{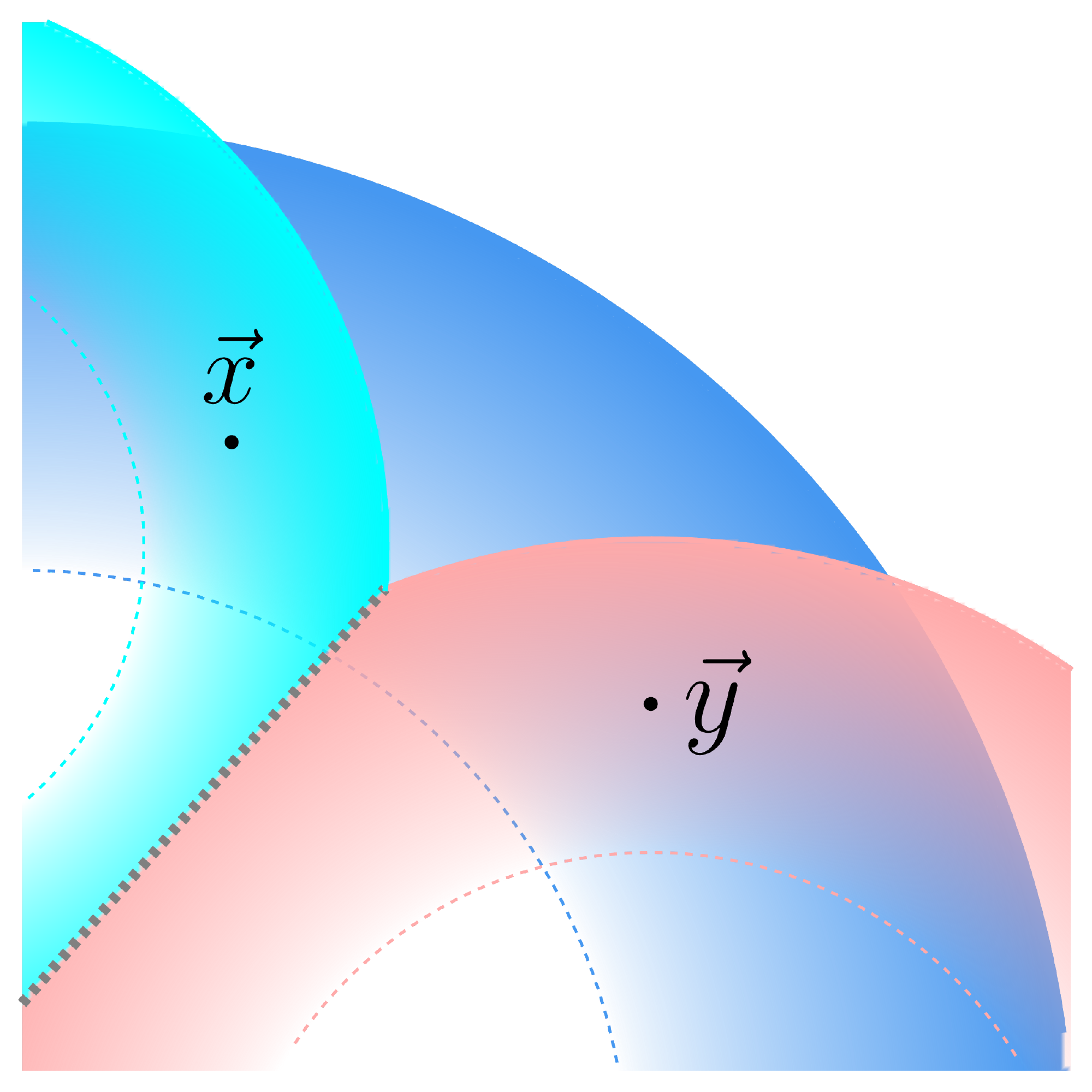}
    \caption{A schematic illustration for the single-shell and double-shell contributions to the energy-momentum tensor with their two-point correlation functions coming from $\vec{x}$ and $\vec{y}$ in the same sound shell (blue shell) or in different sound shells (cyan and red shells).}
    \label{fig:single_double_bubble_app}
\end{figure}

Similar to Ref.~\cite{Jinno:2016vai}, the total power spectrum would receive single-shell and double-shell contributions, depending on whether the sound shell(s) passing through $x$ and $y$ are from the same bubble or two different bubbles, respectively, as shown in Fig.~\ref{fig:single_double_bubble_app}.
At first sight, the single-shell contribution might seem to be zero due to the spherical configuration of the nucleated bubbles, however, it does contribute to the power spectrum and even dominates over the double-shell contribution as explained in the appendix B of Ref.~\cite{Jinno:2016vai}.

\section{Notations}

\begin{figure*}
    \centering
    \subfigure[$t_d/r < 1/v_w$]{
        \begin{minipage}{0.30\linewidth}
            \includegraphics[height = 85pt]{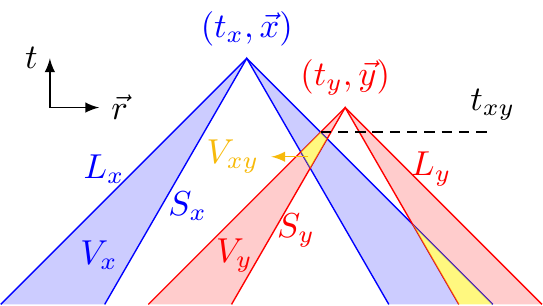} \vspace{20pt}
            \includegraphics[height = 110pt]{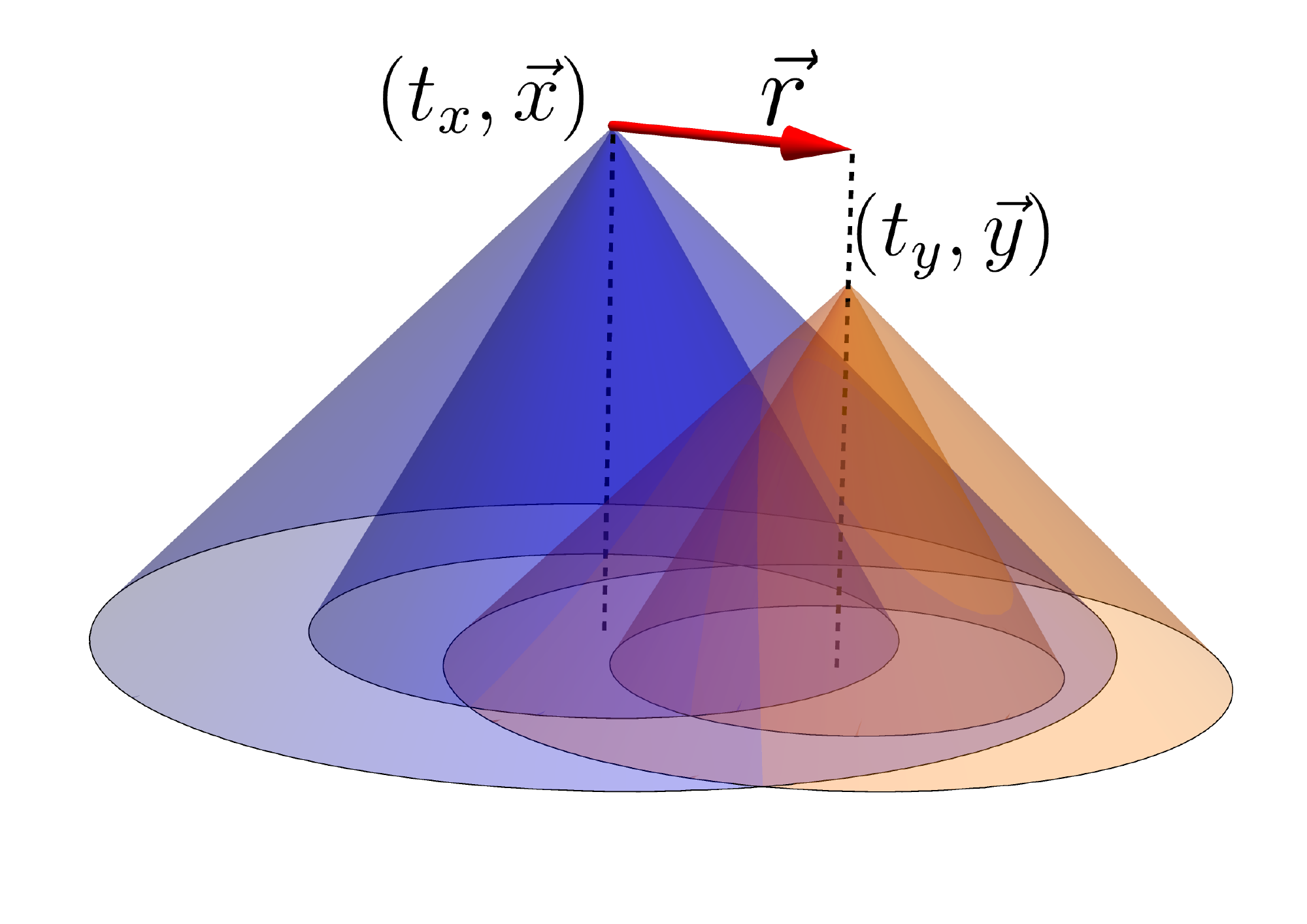}
        \end{minipage}
    }
    \subfigure[$1/v_w < t_d/r <1/c_s$]{
        \begin{minipage}{0.29\linewidth}
            \includegraphics[height = 85pt]{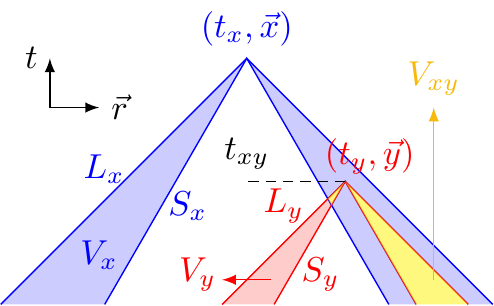} \vspace{20pt}
            \includegraphics[height = 110pt]{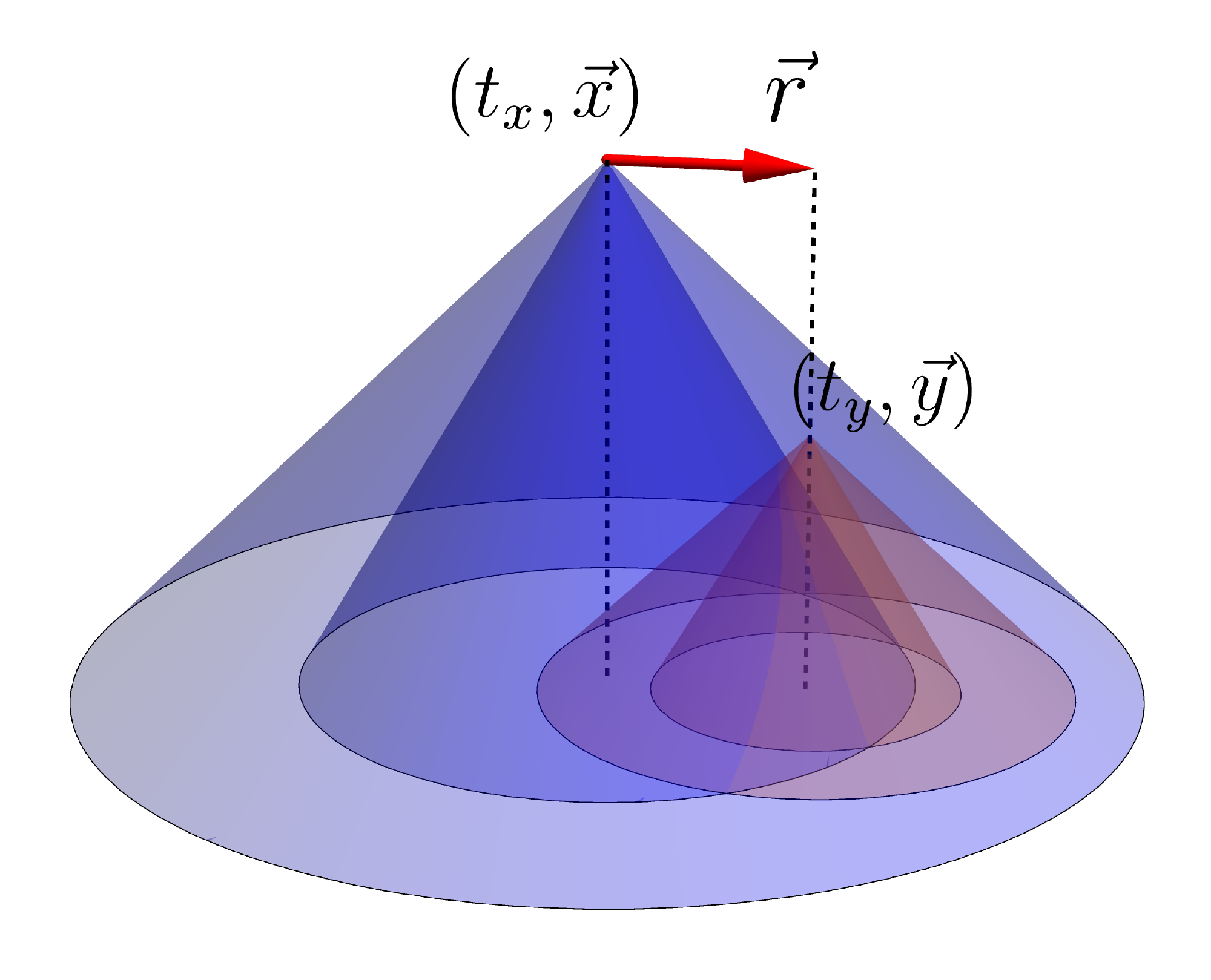}
        \end{minipage}
    }
    \subfigure[$t_d/r > 1 / c_s$]{
        \begin{minipage}{0.29\linewidth}
            \includegraphics[height = 85pt]{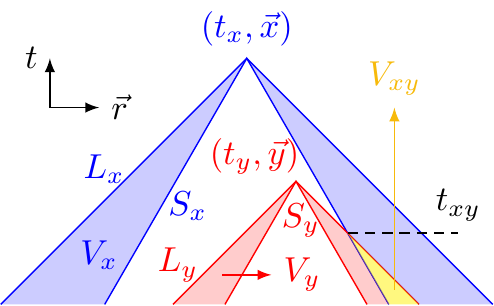} \vspace{20pt}
            \includegraphics[height = 110pt]{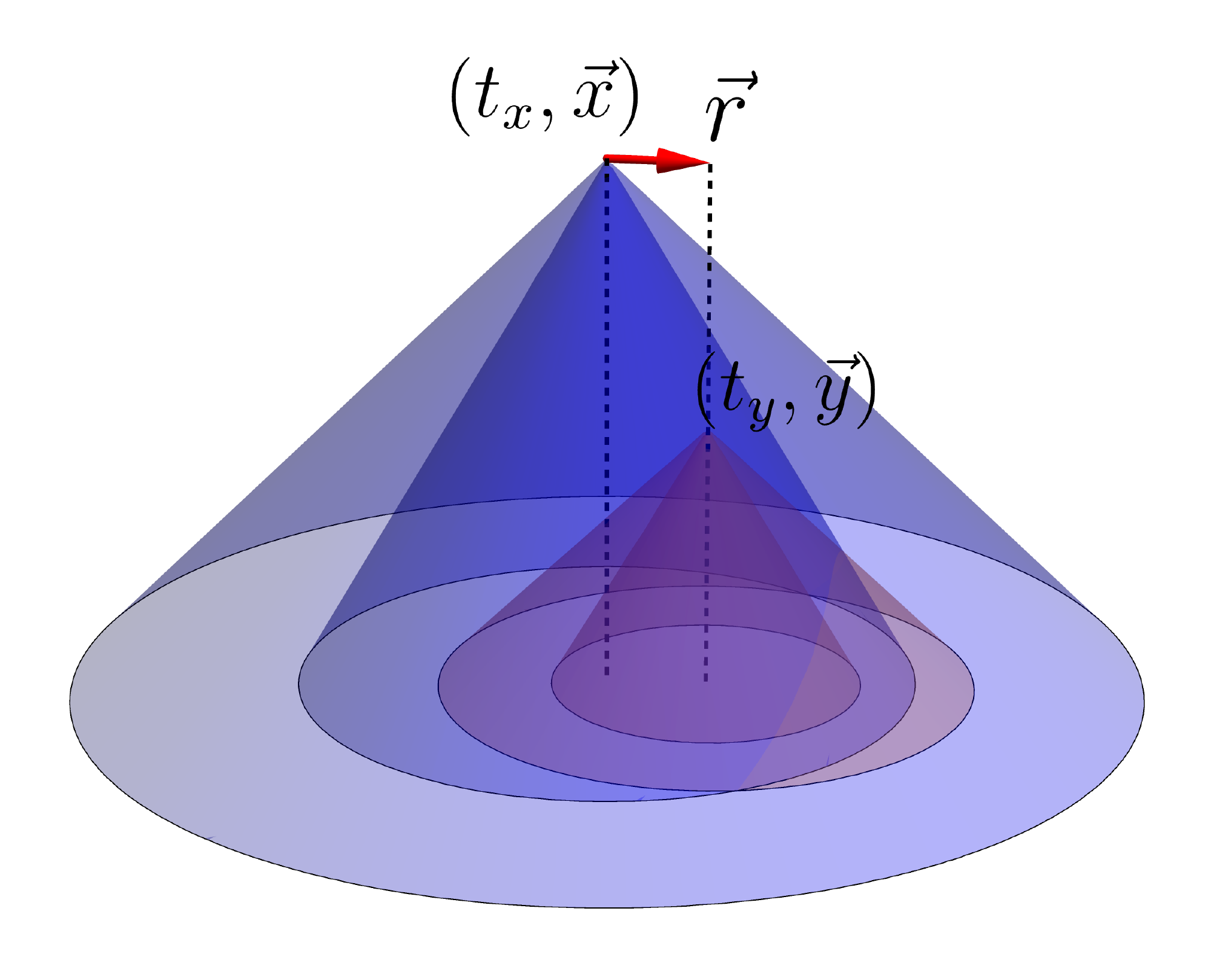}
        \end{minipage}
    }
    \caption{Schematic illustration for the two-point correlator of the energy-momentum tensor from three different configurations shown in  $1+1$ dimensions (top row) and $2+1$ dimensions (bottom row). For the single-shell case, there should be only one bubble nucleated in the region shaded in yellow.}
    \label{fig:three_configurations}
\end{figure*}

\begin{figure}
    \centering
    \includegraphics[width = 0.4\textwidth]{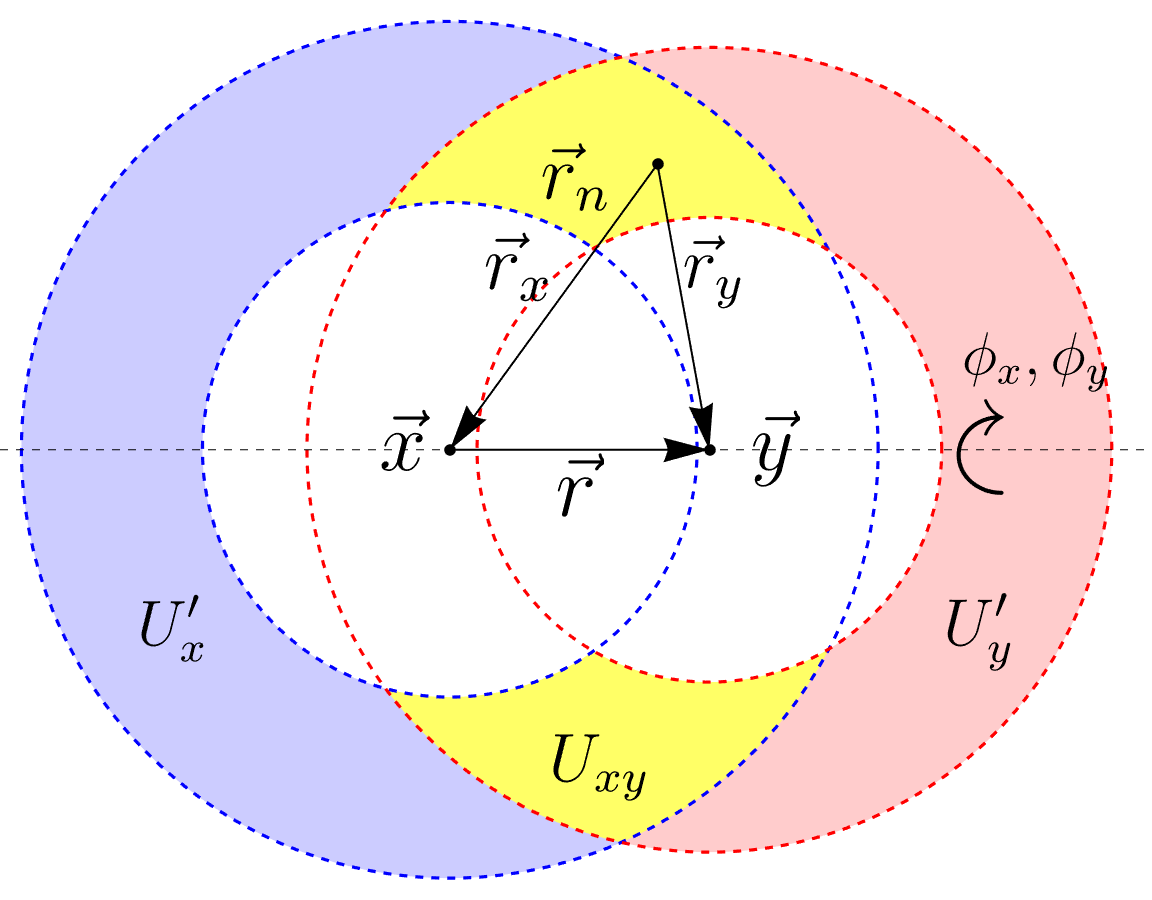}
    \caption{A typical 2-dimensional view for the 3-dimensional region of an equal-time hypersurface. The single-shell contribution comes from a bubble nucleated in the region $U_{xy}$ shaded in yellow, while the double-shell contribution comes from two bubbles nucleated separately in $U_x'$ and $U_y'$ shaded in blue and red. The system has a $SO(2)$ symmetry and is invariant under rotations around $\vec{r}$ direction.}
    \label{fig:3d_scheme}
\end{figure}

To evaluate the two-point correlator~\eqref{eq:stress_correlator} from two space-time points $x=(t_x,\vec{x})$ and $y=(t_y,\vec{y})$, we first define three useful variables as 
\begin{align}
    T = \frac{t_x + t_y}{2}, \quad t_d = t_x - t_y, \quad \vec{r} = \vec{x} - \vec{y}.
\end{align}
Without loss of generality, we might as well consider $t_d \geq 0$. 
We then denote the past $v_w$-cone and past sound cone of a space-time point $x$ as $L_x$ and $S_x$, respectively, and the interiors of them as $V^L_x$ and $V^S_x$, respectively. The region that belongs to the interior of $L_x$ but the exterior of $S_x$ is called $V_x$, and the intersection of $V_x$ and $V_y$ is called $V_{xy}$. There are three possible configurations of two space-time points $x$ and $y$ depending on the ratio $t_d / r$ with $r\equiv |\vec{r}|$, that is,
\begin{enumerate}
    \item For $t_d/r < 1/v_w$, $y$ lives outside $V^L_x$;
    \item For $1/v_w < t_d/r <1/c_s $, $y$ lives inside $V^L_x$ but outside $V^S_x$, thus in $V_x$;
    \item For $t_d/r > 1 / c_s $, $y$ lives inside $V^S_x$.
\end{enumerate}
These three configurations are shown in Fig.~\ref{fig:three_configurations} in $1+1$ dimensions and $2+1$ dimensions respectively. As evident from Fig.~\ref{fig:three_configurations}, the critical time $t_{xy}$ can be expressed as a function of $t_d/r$ for three different configurations as
\begin{align}
    t_{xy} =  \left\{ 
    \begin{aligned}
        & T - \frac{1}{2}\frac{r}{v_w},  & 0\leq \frac{t_d}{r} < \frac{1}{v_w}, \\  
        & T - \frac{t_d}{2} ,  & \frac{1}{v_w} \leq \frac{t_d}{r} < \frac{1}{c_s}, \\
        & T - \frac{v_w + c_s}{v_w - c_s} \frac{t_d}{2} + \frac{r}{v_w - c_s},  & \frac{t_d}{r} \geq \frac{1}{c_s}.
    \end{aligned}
    \right.\label{eq:txy}
\end{align}

The equal-time hypersurface at time $t$ is denoted as $\Sigma_t$, whose intersection with $V_{xy}$ is a $3$-dimensional region called $U_{xy}(t)$. This region only appears when the time $t$ is early enough to intersect with both $V_x$ and $V_y$, namely, $t<t_{xy}$, where the critical time $t_{xy}$ depends on $t_x$, $t_y$, and $r$. Similarly, the projections of $V_x$ ($V_x^L$) and $V_y$ ($V_y^L$) on $\Sigma_t$ are called $U_x$ ($U_x^L$) and $U_y$ ($U_y^L$), respectively. A typical 2-dimensional view for this 3-dimensional region $\Sigma_t$ is shown in Fig.~\ref{fig:3d_scheme}. Consider an arbitrary point $\vec{r}_n\in U_x^L \cup U_y^L$, we define two vectors $\vec{r}_x=\vec{x} - \vec{r}_n$ and $\vec{r}_y=\vec{y} - \vec{r}_n$, whose directions are given by $\vec{n}_x$ and $\vec{n}_y$ in terms of the azimuthal and polar angles around $\vec{r}$ as
\begin{align}
    \begin{aligned}
        \vec{n}_x & = \frac{\vec{x} - \vec{r}_n}{|\vec{x} - \vec{r}_n|} = - (\sin\theta_x \cos\phi_x, \sin\theta_x \sin\phi_x, \cos\theta_x), \\ 
        \vec{n}_y & = \frac{\vec{y} - \vec{r}_n}{|\vec{y} - \vec{r}_n|} = - (\sin\theta_y \cos\phi_y, \sin\theta_y \sin\phi_y, \cos\theta_y).
    \end{aligned}
\end{align}
The azimuthal angles share the same value $\phi_x = \phi_y$. By defining $r_x = |\vec{x} - \vec{r}_n|$ and $r_y = |\vec{y} - \vec{r}_n|$, it is easy to evaluate the polar angles by
\begin{align}
    \begin{aligned}
        \cos\theta_x & = - \frac{r_x^2 + r^2 - r_y^2}{2 r ~ r_x}, \quad
        \cos\theta_y & = \frac{r_y^2 + r^2 - r_x^2}{2 r ~ r_y}.
    \end{aligned}
\end{align}

\section{Evaluation}

 We next turn to explicitly evaluate the time-independent part of the GW power spectrum,
\begin{align}
    \Delta(k) &=  k^3 \int_{t_i}^{t_f}\md t_x \int_{t_i}^{t_f} \md t_y \cos(k(t_x - t_y)) \int \md^3 r e^{i \vec{k} \cdot \vec{r}}  \nonumber \\
    & \quad\quad \times \Lambda_{ij,kl}(\hat{k}) \Lambda_{ij,mn}(\hat{k})  \langle T_{kl}(t_x,\vec{x})T_{mn}(t_y,\vec{y})\rangle, \label{eq:power_propto}
\end{align}
where the two-point correlators of the energy-momentum tensor are evaluated differently for the single-shell and double-shell cases. For the single-shell contribution, the two-point correlator with a superscript $(s)$ reads
\begin{align}
    & \langle T_{ij}(t_x,\vec{x})T_{kl}(t_y,\vec{y})\rangle^{(s)} = P(t_x,t_y,r) \nonumber \\
    & \quad \times \int_{-\infty}^{t_{xy}} \md t_n \Gamma(t_n) \int_{U_{xy}}\md^3 r_n \tilde{T}_{ij,kl}^{(s)}(t_n,t_x,t_y,\vec{r}), \label{eq:single_TT}
\end{align}
where $\tilde{T}_{ij,kl}^{(s)}(t_n,t_x,t_y,\vec{r})$ is the value of $T_{ij}(x)T_{kl}(y)$ from the same bubble nucleated at time $t_n$ and volume element $\mathrm{d}^3r_n$ in $U_{xy}$ for a given nucleation rate $\Gamma(t)$, and $P(t_x,t_y,r)$ is the probability for both $x$ and $y$ staying in the false vacuum to ensures that no other bubbles ever reach $x$ and $y$ before the bubble we consider. For the double-shell case, the two-point correlator with a superscript $(d)$ reads
\begin{align}
    & \langle T_{ij}(t_x,\vec{x})T_{kl}(t_y,\vec{y})\rangle^{(d)} =  P(t_x,t_y,r) \nonumber \\
    & \quad \times \int_{-\infty}^{t_{xy}} \md t_{xn} \Gamma(t_{xn}) \int_{U_x'}\md^3 r_{xn} \tilde{T}_{x,ij}^{(d)}(t_{xn},\vec{x}_n;t_x,\vec{r}) \nonumber \\
    & \quad \times \int_{-\infty}^{t_{xy}} \md t_{yn} \Gamma(t_{yn}) \int_{U_y'}\md^3 r_{yn} \tilde{T}_{y,kl}^{(d)}(t_{yn},\vec{y}_n;t_y,\vec{r}),
    \label{eq:double_TT}
\end{align}
where $\tilde{T}_{p,ij}(t_{pn},\vec{p}_n;t_p,\vec{r})$ is the value of $T_{ij}(p)$ by a bubble nucleated at time $t_{pn}$ in $U_p'$ with $p=x,y$. The region $U_p'$ is defined as the set of spatial points $q$ satisfying $q\in U_p$ and $q \notin U_{xy}^L$. $\Gamma(t)$ and $P(t_x,t_y,r)$ are the same as those in the single-shell case.

As we have assumed in the main context, we will focus on the exponential decay rate $\Gamma(t)=\Gamma(t_*)\exp[\beta(t-t_*)]$ with the phase transition completed roughly around $t_*$. Then, similar to the case in Ref.~\cite{Jinno:2016vai} to evaluate the probability $P(t_x,t_y,r)$ for two points with space-like separation $ r > t_d$ staying in the false vacuum when bubbles could expand approximately with the speed of light, our evaluation on the probability $P(t_x,t_y,r)$ with a finite bubble wall velocity $v_w$ arrives at
\begin{align}
    P(t_x, t_y ,r) & = e^{-I(T,t_d,r)},
    \label{eq:Possibility_in_FV} \\
    I(T,t_d,r) & = 8\pi v_w^3 \frac{\Gamma(T)}{\beta^4} \mathcal{I}\left(\beta t_d,\beta \frac{r}{v_w}\right),
\end{align}
where $t_x$ and $t_y$ are encoded in their linear combinations $T$ and $t_d$. $\mathcal{I}(t,r)$ is a dimensionless function  given by,
\begin{align}
    \mathcal{I}(t,r) & = \left\{ 
    \begin{aligned}
        & e^{-t/2} + e^{t/2} + \frac{t^2 -r^2 - 4r}{4r}e^{-r/2}, & t \leq r, \\
        & e^{-t/2}, & t>r.
    \end{aligned}
    \right.
\end{align}

Now we can plug Eq.~\eqref{eq:Possibility_in_FV} into \eqref{eq:power_propto} with a conversion of the integral variables from $\{t_x, t_y\}$ to $\{T, t_d\}$, and then take the limit $t_i \to - \infty$, $t_f \to \infty$ to obtain
\begin{align}
    \Delta(k) = & ~ 2 k^3\int_0^{+\infty} \md t_d \int_{-\infty}^{+\infty} \md T P(t_x, t_y, r) \nonumber \\
    & \quad \times (\mathrm{some~functions~of~}t_x, t_y, k). \label{eq:Delta_Int_T}
\end{align}
In the second line, the integral takes different forms for single-shell and double-shell cases. The time dependence of the energy-momentum tensor tells us that they are functions of $t-t_n$, thus $\tilde{T}_{ij,kl}^{(s)}(t_n,t_x,t_y,\vec{r})$ and $\tilde{T}_{p,ij}(t_{pn},\vec{p}_n;t_p,\vec{r})$ are functions of $T-t_n$ and $t_d$. Similar analyses hold for the integral regions $U_{xy}$ and $U_p'$. Rewrite the integrals over nucleation time in Eq.~\eqref{eq:single_TT} and \eqref{eq:double_TT}, one could find that the $T$ dependence in the time integrals can be summarized in a simple factor $\Gamma(T)$. Now the two-point correlator for the single-shell case reads
\begin{align}
    & \langle T_{ij}(t_x,\vec{x})T_{kl}(t_y,\vec{y})\rangle^{(s)} / P(t_x, t_y ,r) \nonumber \\
    & \quad = \int_{-\infty}^{t_{xy}} \md t_n \Gamma(t_n)  \int_{U_{xy}}\md^3 r_n \tilde{T}_{ij,kl}^{(s)}(t_n,t_x,t_y,\vec{r}) \nonumber \\
    & \quad = \int_{-\infty}^{t_{xy}} \md (t_n - T) \Gamma(T) e^{\beta (t_n - T)}   \nonumber \\
    & \quad \quad \quad \times \int_{U_{xy}(T-t_n,t_d,\vec{r})}\md^3 r_n \tilde{T}_{ij,kl}^{(s)}(T-t_n,t_d,\vec{r}_n,\vec{r}) \nonumber \\
    & \quad = \Gamma(T)\int_{-\infty}^{t_{xy}-T}\md u ~ e^{\beta u} \int_{U_{xy}}\md^3 r_n \tilde{T}_{ij,kl}^{(s)}(u,t_d,\vec{r}_n,\vec{r}),
    \label{eq:TTbyP_single}
\end{align}
where $u\equiv t_n - T$. Here we use the fact that $t_{xy} - T$ is actually $T$ independent as seen from Eq.~\eqref{eq:txy}. Thus, the only $T$ dependence is in the $\Gamma(T)$ factor. The two-point correlator for the double-shell case reads similarly
\begin{align}
    & \langle T_{ij}(t_x,\vec{x})T_{kl}(t_y,\vec{y})\rangle^{(d)} /  P(t_x,t_y,r) \nonumber \\
    & \quad = \Gamma(T)^2 \iint \md^4 x_n\md^4 y_n\times (\mathrm{functions~of~}x_n,y_n,x,y).
    \label{eq:TTbyP_double}
\end{align}
It is worth noting that $T$ only marks the ``average time'' of the system, while $t_d$ and $r$ together describe the relative position of the two space-time points $x$ and $y$. Hence, the correlation information of the system is contained in $t_d$ and $r$ but not $T$.

Here we provide a formula for the integral over $T$ in Eq.~\eqref{eq:Delta_Int_T} as
\begin{align}
    & \int_{-\infty}^\infty \md T ~\Gamma(T)^l \exp\left( - 8\pi v_w^3 \frac{\Gamma(T)}{\beta^4} \mathcal{I}\left(\beta t_d,\beta \frac{r}{v_w}\right)\right) \nonumber \\
    & = \Gamma(t_*)^l \int_{-\infty}^\infty \md T \exp\left( l\beta(T-t_*) - C \Gamma(t_*)e^{\beta(T-t_*)}\right) \nonumber \\
    & = \frac{\Gamma(t_*)^l}{\beta} \frac{(l-1)!}{(C \Gamma(t_*))^l} = \frac{1}{\beta}\left( 8\pi \frac{v_w^3}{\beta^4} \mathcal{I}\left(\beta t_d,\beta \frac{r}{v_w}\right) \right)^{-l},
    \label{eq:T_integral}
\end{align}
where $l=1,2$ for single-shell and double-shell respectively, and $C \equiv 8\pi v_w^3\beta^{-4} \mathcal{I}\left(\beta t_d,\beta r/v_w\right)$.

\section{Single-shell contribution}

Now we derive the analytic expression for the single-shell spectrum $\Delta^{(s)}$. Since there is no characteristic direction other than $\vec{r}$, we first rewrite the two-point correlator $\langle T_{ij}T_{kl}\rangle^{(s)}$ in a form as
\begin{align}
    \langle T_{ij}T_{kl}\rangle^{(s)} =& a_1 \delta_{ij}\delta_{kl} + a_2 \delta_{i\left(k\right.}\delta_{\left.l\right)j} + b_1 \delta_{ij}\hat{r}_k \hat{r}_l + b_2 \delta_{kl}\hat{r}_i \hat{r}_j \nonumber \\
    & + \frac{b_3}{2}\left( \delta_{i\left(k\right.}\hat{r}_{\left.l\right)} \hat{r}_j + \delta_{j\left(k\right.}\hat{r}_{\left.l\right)} \hat{r}_i \right) + c_1 \hat{r}_i \hat{r}_j \hat{r}_k \hat{r}_l.
\end{align}
Then, we project this two-point correlator by $\Lambda_{ij,kl}(\hat{k})$ in the transverse-traceless gauge with $\cos\alpha_{rk} \equiv \hat{r}\cdot\hat{k}$ as
\begin{align}
    & \Lambda_{ij,kl}(\hat{k})\Lambda_{ij,mn}(\hat{k}) \langle T_{kl}T_{mn}\rangle^{(s)} \nonumber \\
    & \quad = 2 a_2 + (1 - \cos^2 \alpha_{rk}) b_3 + \frac{1}{2}(1 - \cos^2 \alpha_{rk})^2 c_1.\label{eq:LLTT_single}
\end{align}
Regarding that $a_2$, $b_3$ and $c_1$ are independent of $\hat{k}$, we can plug \eqref{eq:LLTT_single} into \eqref{eq:power_propto} and perform the angular parts of the spatial integral. Using the formulas concerning with the spherical Bessel functions $j_n(x)$,
\begin{align}
    & \int_{-1}^1 \md c~  e^{icx} = 2 j_0(x), \quad \int_{-1}^1 \md c~  e^{icx} (1-c^2) = 4 \frac{j_1(x)}{x}, \nonumber\\
    & \int_{-1}^1 \md c~  e^{icx} (1-c^2)^2 = 16 \frac{j_2(x)}{x^2}, \label{eq:SphericalJ}
\end{align}
we can directly obtain the single-shell spectrum as
\begin{align}
\Delta^{(s)}&(k) =  2 k^3\int_0^{+\infty} \md t_d \int_{-\infty}^{+\infty} \md T \cos(k t_d) \int_0^{+\infty} \md r\,r^2 \nonumber \\
& \times 2\pi \left(4 j_0(k r) a_2 + 4 \frac{j_1(k r)}{k r} b_3 + 8 \frac{j_2(k r)}{(k r )^2} c_1 \right).
    \label{eq:deltas_int3}
\end{align}

We next evaluate $a_2$, $b_3$ and $c_1$. By identifying $\hat{z}$ direction as $\hat{r}$, the covariant parameters $a_2$, $b_3$ and $c_1$ can be expressed in terms of spatial components of the correlator as
\begin{align}
    a_2 = & \langle T_{12}T_{12}\rangle^{(s)}, \nonumber \\
    b_3 = & 4 \left( \langle T_{13}T_{13}\rangle^{(s)} - \langle T_{12}T_{12}\rangle^{(s)}\right), \nonumber \\
    c_1 = & \langle T_{11}T_{11}\rangle^{(s)} - \left( \langle T_{11}T_{33}\rangle^{(s)} + \langle T_{33}T_{11}\rangle^{(s)}\right) \nonumber\\
    & - 4 \langle T_{13}T_{13}\rangle^{(s)} + \langle T_{33}T_{33}\rangle^{(s)},
    \label{eq:abc}
\end{align}
where the subscript $3$ denotes the $\hat{z}$ direction ($\hat{r}$ direction) and the subscripts $1$ and $2$ denote two equivalent directions $\hat{x}$, $\hat{y}$ which are perpendicular to $\hat{z}$ and to each other.

Since $r$ serves as a length scale of the system, the integrals of Eq.~\eqref{eq:TTbyP_single} could be rescaled by $r$ as
\begin{align}
    & \langle T_{ij}(t_x,\vec{x})T_{kl}(t_y,\vec{y})\rangle^{(s)} \left/ \left( P(t_x, t_y ,r) \Gamma(T) r^4 \right) \right. \nonumber \\
    & = \int_{-\infty}^{t_{xy}-T}\md\left(\frac{u}{r}\right) ~ e^{\beta r \frac{u}{r}} \int_{U_{xy}}\md^3 \left(\frac{r_n}{r}\right)~ \tilde{T}_{ij,kl}^{(s)}(u,t_d,\vec{r}_n,\vec{r}).
    \label{eq:integral_tnrn_single}
\end{align}
From Eq.~\eqref{eq:stress_tensor_app} one could see that the dependence on $r$ and $t$ in $\tilde{T}_{ij,kl}^{(s)}(u,t_d,\vec{r}_n,\vec{r})$ are not independent but in a combination $t/r$, Thus, the integrals above can be done with respect to dimensionless variables $u/r$ and $r_n/r$. For later convenience we define
\begin{align}
    \begin{aligned}
        a_2 = P(t_x, t_y ,r) \Gamma(T) r^4 F_a(t_d /r), \\
        b_3 = P(t_x, t_y ,r) \Gamma(T) r^4 F_b(t_d /r), \\
        c_1 = P(t_x, t_y ,r) \Gamma(T) r^4 F_c(t_d /r),
    \end{aligned}
    \label{eq:abc_to_F}
\end{align}
where $F_a$, $F_b$, and $F_c$ are the integrals of the linear combinations of spatial components of the energy-momentum tensor given by Eq.~\eqref{eq:abc} and \eqref{eq:integral_tnrn_single}. Replacing $a_2$, $b_3$, and $c_1$ in Eq.~\eqref{eq:deltas_int3} with \eqref{eq:abc_to_F} and performing the integral over $T$, we finally arrive at the single-shell spectrum,
\begin{align}
    & \Delta^{(s)}(k)  \nonumber \\
    & =  2 v_w^{-3} \beta^3 k^3\int_0^{+\infty} \md t_d \int_0^{+\infty} \md r \frac{\cos(k t_d)}{\mathcal{I}(\beta t_d,\beta r / v_w)} \nonumber \\
    & \quad \times r^6 \left(j_0(k r) F_a + \frac{j_1(k r)}{k r} F_b + 2 \frac{j_2(k r)}{(k r )^2} F_c \right) \nonumber\\
    & = \frac{2}{v_w^3}\beta^{-2} \tilde{k}^3
    \int_0^{+\infty} \md \tilde{t} \int_0^{+\infty} \md \tilde{r} \frac{\cos(\tilde{k}\tilde{t})}{\mathcal{I}(\tilde{t},\tilde{r}/ v_w)} \nonumber\\
    & \quad \times \tilde{r}^6 \left(j_0(\tilde{k} \tilde{r}) F_a + \frac{j_1(\tilde{k} \tilde{r})}{\tilde{k} \tilde{r}} F_b + 2 \frac{j_2(\tilde{k} \tilde{r})}{(\tilde{k} \tilde{r})^2} F_c \right)
    \label{eq:delta_single_int2}
\end{align}
where the variables $\tilde{k} = k/\beta$, $\tilde{t} = \beta t_d$, and $\tilde{r} = \beta r$ are rescaled by $\beta$. Note that $F_{a,b,c}$ shares the same dimension as the enthalpy square and can be rescaled by the asymptotic enthalpy at null infinity as $\tilde{F}_{a,b,c} = F_{a,b,c}/w_N^2$.
Similar dimensional analysis also holds for the double-shell spectrum which we will pursuit below.

\section{Double-shell contribution}

Next, we evaluate the double-shell spectrum. Following similar analyses, the correlator \eqref{eq:double_TT} can be decomposed as
\begin{align}
    \langle T_{ij}T_{kl}\rangle^{(d)}  = & P(t_x, t_y ,r) \nonumber\\
    & ~ \times (A_x(T,t_d,r)\delta_{ij} + B_x(T,t_d,r) \hat{r}_i \hat{r}_j) 
    \nonumber \\
    & \quad \times (A_y(T,t_d,r)\delta_{kl} + B_y(T,t_d,r) \hat{r}_k \hat{r}_l).
\end{align}
By identifying $\hat{r}$ direction as the $\hat{z}$ direction, $A_p(T,t_d,r)$ and $B_p(T,t_d,r)$ can be expressed in terms of the integrals of energy-momentum tensors. For example, $B_p(T,t_d,r)$ is given by
\begin{align}
    & B_p(T,t_d,r)  \nonumber\\ 
    &= \int_{-\infty}^{t_{xy}} \md t_{pn} \Gamma(t_{pn}) \int_{U_p'}\md^3 r_{pn} \left(\tilde{T}_{p,33}^{(d)} - \tilde{T}_{p,11}^{(d)} \right) \nonumber \\
    &=  r^4 \Gamma(T) \int_{-\infty}^{t_{xy}-T}\md\left(\frac{u}{r}\right) ~ e^{\beta r \frac{u}{r}}~ \nonumber \\
    & \quad \times   \int_{U_{p}'}\md^3 \left(\frac{r_{pn}}{r}\right) \left(\tilde{T}_{p,33}^{(d)} - \tilde{T}_{p,11}^{(d)} \right)\nonumber\\
    &\equiv r^4 \Gamma(T) G_p(t_d/r)
\end{align}
for $p = x,y$. After projection in the transverse-traceless gauge, there is only one non-vanishing term,
\begin{align}
    & \Lambda_{ij,kl}(\hat{k})\Lambda_{ij,mn}(\hat{k}) \langle T_{kl}T_{mn}\rangle^{(d)} \nonumber \\
    &  =  \frac{1}{2}(1 - \cos^2 \alpha_{rk})^2 P(t_x, t_y ,r)  B_x(T,t_d,r) B_y(T,t_d,r) \nonumber\\ 
    &  = \frac{1}{2}(1 - \cos^2 \alpha_{rk})^2 P(t_x, t_y ,r) \Gamma(T)^2  r^8 G_x(t_d/r)G_y(t_d/r). \label{eq:LLTT_double}
\end{align}
Using Eq.~\eqref{eq:T_integral} and \eqref{eq:SphericalJ} for performing the integrals over $T$ and angular part of $\vec{r}$, we can trade the factor $P(t_x, t_y ,r) \Gamma(T)^2$ and $\frac{1}{2}(1 - \cos^2 \alpha_{rk})^2$ to the already known factors,
\begin{align}
    & P(t_x, t_y ,r) \Gamma(T)^2 \to \frac{\beta^7}{16\pi^2 v_w^6} \mathcal{I}\left(\beta t_d,\beta \frac{r}{v_w}\right)^{-2} , \\
    & \frac{1}{2}(1 - \cos^2 \alpha_{rk})^2 e^{ikr\cos\alpha_{rk}} \to 8\frac{j_2(kr)}{(kr)^2}.
\end{align}

It is worth noting that there is no contribution to the double-shell spectrum from $t_d/r > 1/v_w$ since only one bubble is required to be nucleated within $V^L_x$. The integral over $t_d$ is thus in the interval $(0,r/v_w)$ instead of $(0,+\infty)$ any more. Finally, the double-shell spectrum reads
\begin{align}
    & \Delta(k)^{(d)} \nonumber\\
    & =  \frac{1}{2\pi} v_w^{-6} \beta^7 k^3\int_0^{r/v_w} \md t_d \int_0^{+\infty} \md r  \frac{\cos(k t_d)}{\mathcal{I}(\beta t_d,\beta r / v_w)^2} \nonumber \\
    &  \quad \times r^{10}  \frac{j_2(k r)}{(k r )^2} G_x\left(t_d/r\right) G_y\left( t_d/r \right) \nonumber \\
    & =  \frac{1}{2\pi v_w^6} \beta^{-2} \tilde{k}^3
    \int_0^{1/v_w} \md \tilde{t} \int_0^{+\infty} \md \tilde{r} \frac{\cos(\tilde{k}\tilde{t})}{\mathcal{I}(\tilde{t},\tilde{r}/ v_w)^2} \nonumber\\
    & \quad \times \tilde{r}^{10} \left( \frac{j_2(\tilde{k} \tilde{r})}{(\tilde{k} \tilde{r})^2} G_x\left(\tilde{t}/\tilde{r}\right) G_y\left( \tilde{t}/\tilde{r} \right) \right)
    \label{eq:delta_double_int2}
\end{align}

\section{Numerical evaluations}

\begin{figure*}
    \centering
    \subfigure{
        \includegraphics[width = 0.45\textwidth]{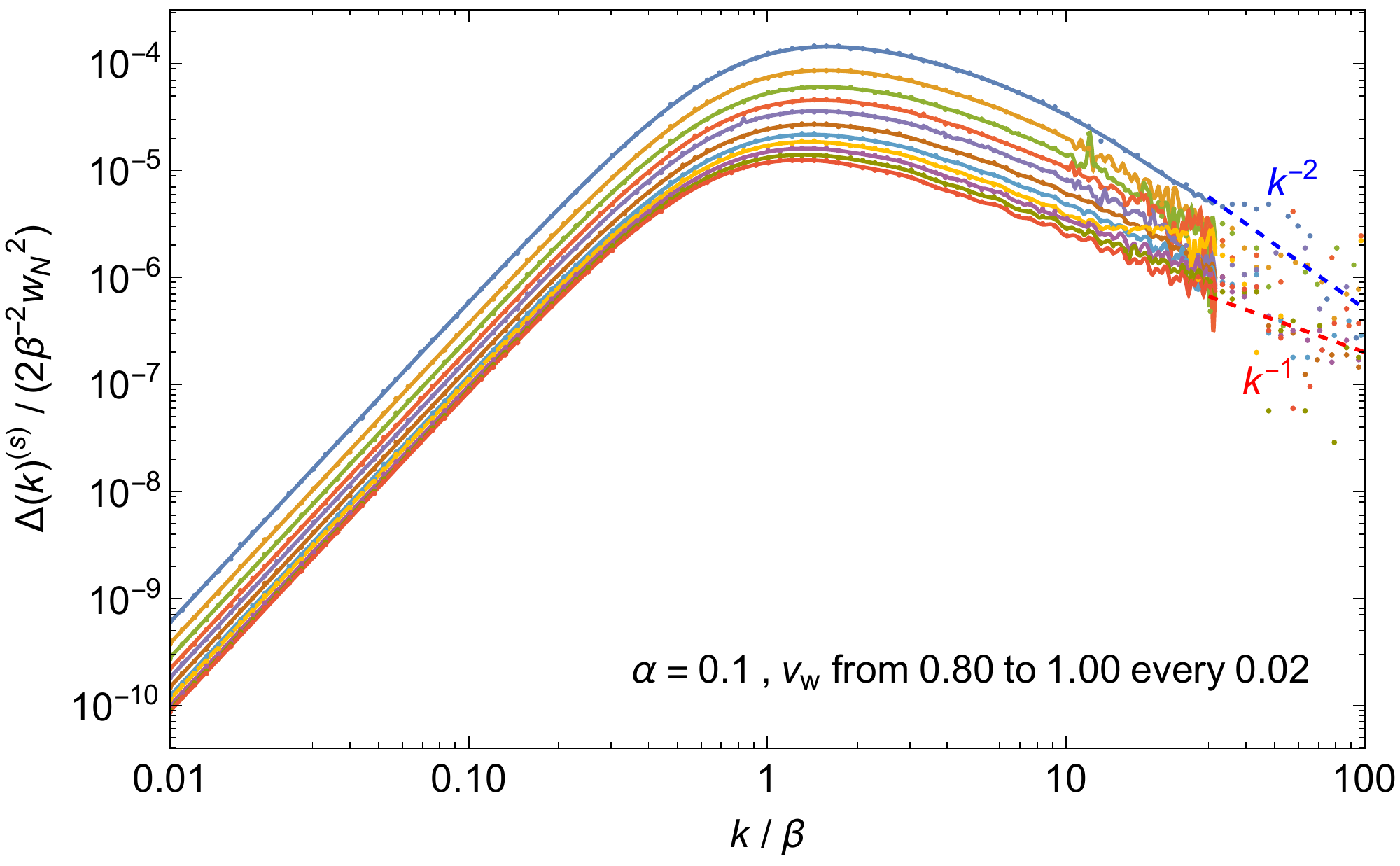}
    }
    \subfigure{
        \includegraphics[width = 0.45\textwidth]{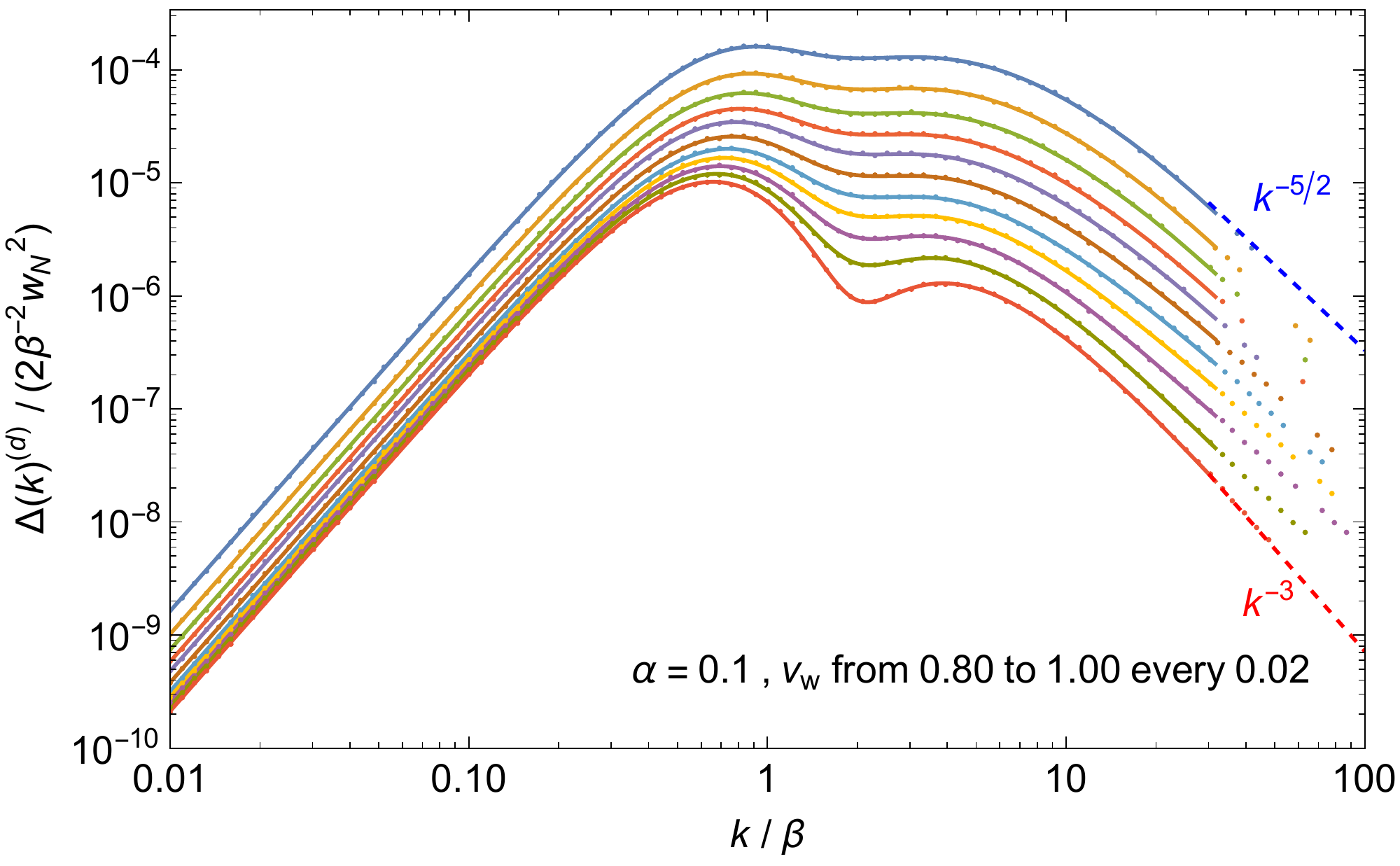}
    }
    \caption{The single-shell (left) and double-shell (right) power spectra for $\alpha=0.1$. The discrete points are the numerical results and solid lines are their interpolating functions. The curves correspond to different $v_w$ from $0.80$ to $1.00$ with interval $\Delta v_w = 0.02$ from the top one to the bottom one, respectively. The blue and red dashed lines are the asymptotic behaviors of the power spectra with $v_w=0.8$ and $v_w=1$, respectively. All these power spectra scale as $k^3$ at low frequencies.} \label{fig:Power_a0.1_of_v}
\end{figure*}

\begin{figure*}
    \centering
    \subfigure{
        \includegraphics[width = 0.95\textwidth]{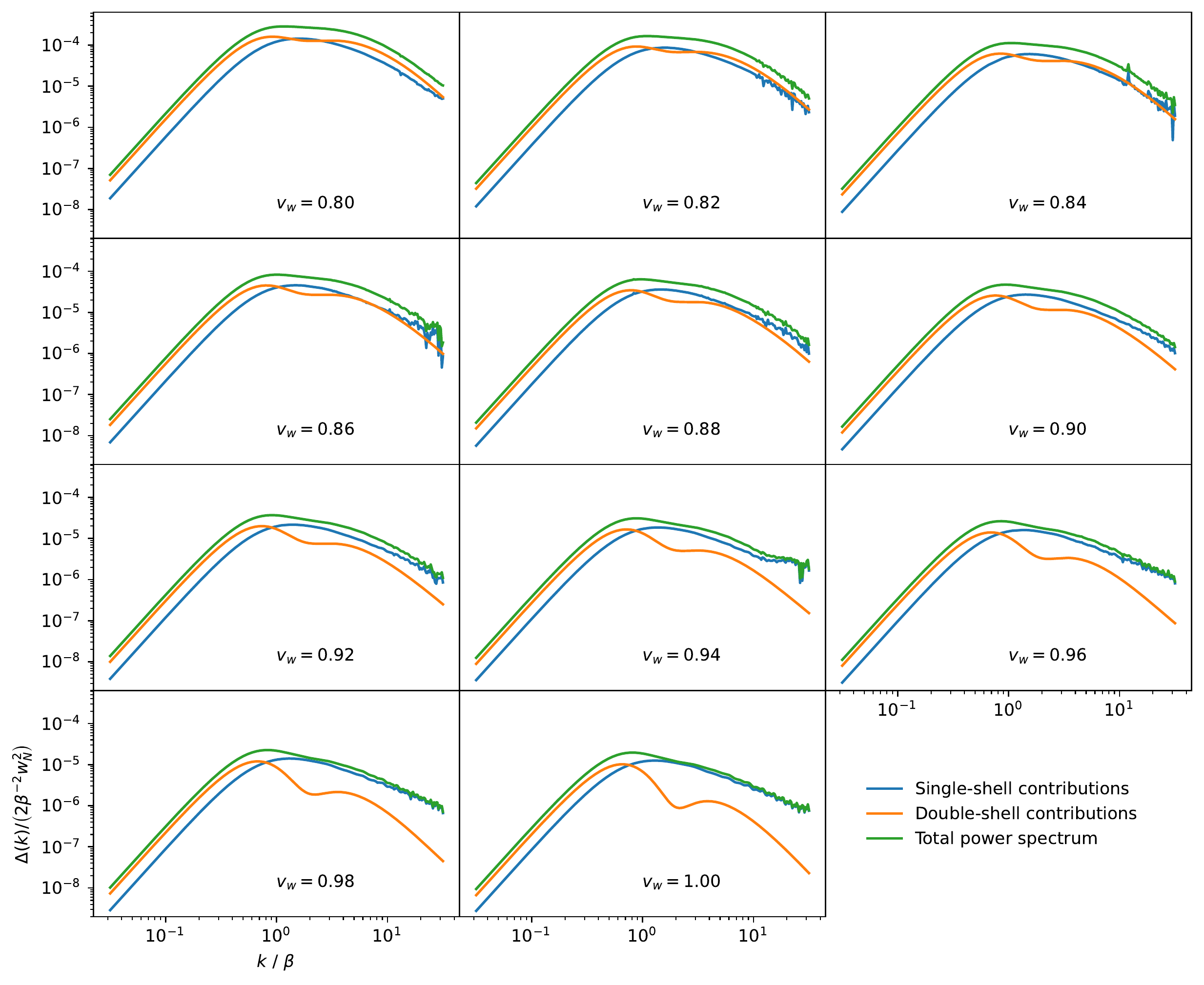}
    }
    \caption{The full shapes of power spectra from the single-shell (blue) and double-shell (orange) contributions to the total power spectrum (green) for different $v_w$ with fixed $\alpha = 0.1$.} \label{fig:Power_different_v}
\end{figure*}

With the analytic forms~\eqref{eq:delta_single_int2} and ~\eqref{eq:delta_double_int2}, it is straightforward to evaluate them numerically for some illustrative examples with fixed $\alpha=0.1$ but varying $v_w=0.8\sim1.0$ as shown in Fig.~\ref{fig:Power_a0.1_of_v} and Fig.~\ref{fig:Power_different_v}. It is easy to see the following patterns:
\begin{enumerate}
\item The asymptotic slope in the infrared always approaches a $k^3$-scaling as usually expected from causality for both single-shell and double-shell contributions to the total GW power spectrum;
\item The double-shell contribution always dominates over the single-shell contribution at low frequencies, while the single-shell contribution gradually dominates over the double-shell contribution at high frequencies with an increasing bubble wall velocity;
\item The asymptotic slopes in the ultraviolet vary with an increasing bubble wall velocity from $k^{-2}$ to $k^{-1}$ for the single-shell spectrum and from $k^{-5/2}$ to $k^{-3}$ for the double-shell spectrum.
\end{enumerate}
All these complicated behaviors combined together give rise to a widened dome around the peak frequency with a decreasing detonation wall velocity.

\bibliographystyle{utphys}
\bibliography{ref}

\end{document}